%% file: main.tex
\documentclass{article}

\PassOptionsToPackage{numbers, compress}{natbib}

\usepackage[final,main]{neurips_2026}

\usepackage[american]{babel}

\usepackage[allcolors=NavyBlue,colorlinks=true,backref=page]{hyperref}       %
\usepackage{subcaption}

\usepackage{natbib} %
\usepackage{mathtools} %
\usepackage{booktabs} %
\usepackage{xurl} %

\usepackage{microtype}
\usepackage{nicefrac}
\usepackage{xspace}
\usepackage{amsmath,amssymb,amsthm}
\usepackage{multirow,makecell,threeparttable,diagbox}
\usepackage{wrapfig}
\usepackage[dvipsnames]{xcolor} %
\usepackage{algorithm}
\usepackage[noend]{algpseudocode}
\usepackage{prettyref}
\usepackage{placeins} %
\usepackage{tabularx}
\usepackage{array}
\newcolumntype{L}{>{\raggedright\arraybackslash}X}

\usepackage{enumitem}
\usepackage{adjustbox}

\newif\ifshowtodos
\showtodostrue %

\input{macros}

\title{
Leveraging generative hallucination and biophysics-informed modeling for unified biomolecular sequence–structure co-design}

\author{Xuefeng Liu\textsuperscript{1,2,$\dagger$}\thanks{Equal Contribution and co-Lead. $\dagger$: Part of this work was completed at the University of Chicago.} ,~\textbf{Mingxuan Cao\textsuperscript{3$\ast$}},~\textbf{Xiao Luo\textsuperscript{4$\ast$}},
~\textbf{Songhao Jiang\textsuperscript{2}}\\
~\textbf{Tobin Sosnick\textsuperscript{5}},
~\textbf{Jinbo Xu\textsuperscript{6}},
~\textbf{Louis. J. Maher\textsuperscript{7}},
~\textbf{Rick L. Stevens\textsuperscript{2,8}}\\
\textsuperscript{1}Department of Medicine, University of Florida\\
\textsuperscript{2}Department of Computer Science, University of Chicago\\
\textsuperscript{3}Data Science Institute, University of Chicago\\
\textsuperscript{4}Department of Pathology, University of Chicago\\
\textsuperscript{5}Department of Biochemistry and
Molecular Biology, University of Chicago\\
\textsuperscript{6}Toyota Technological Institute at Chicago\\
\textsuperscript{7}Department of Biochemistry and Molecular Biology, Mayo Clinic\\
\textsuperscript{8}Argonne National Laboratory \\ \\
{Corresponding author}: Xuefeng.Liu@medicine.ufl.edu
}

\begin{document}

\maketitle

\input{abstract}

\input{introduction}

\input{related_works}

\input{algorithm}

\input{experiments}

\input{conclusion}

\bibliography{reference}

\clearpage
\appendix
\section*{Appendix}
\FloatBarrier       %

\input{supp_related_works}

\input{supp_algorithm}
\input{supp_additional_experiments}
\input{supp_preliminaries}

\end{document}

%% file: macros.tex
\newcommand\state{s}

\newcommand\QFunc{\ensuremath{\ensuremath{Q}}}

\newcommand\Num{\ensuremath{N}}

\newcommand\action{\ensuremath{a}}

\newcommand\CUPUCT{\ensuremath{\text{C}\mathcal{U}\text{-PUCT}}}

\newcommand{\algname}{\textsc{MCTH}\xspace}

\newif\iffinal
\finaltrue

\iffinal
    \newcommand{\fix}[1]{#1}
    \newcommand{\XLinline}[1]{}
    \newcommand{\SH}[1]{}
    \newcommand{\SHinline}[1]{}
    \newcommand{\CT}[1]{}
    \newcommand{\CTinline}[1]{}
    \newcommand{\DP}[1]{}
    \newcommand{\DPinline}[1]{}
    \newcommand{\AC}[1]{}
    \newcommand{\ACinline}[1]{}
    \newcommand{\note}[1]{}
    \newcommand{\pref}[1]{}
\else
    \newcommand{\fix}[1]{{\color{red} #1}}
    \newcommand{\SH}[1]{\todo[fancyline,color=Maroon!40]{SH: #1}\xspace}
    \newcommand{\DP}[1]{\todo[fancyline,color=Purple!40]{DP: #1}\xspace}
    \newcommand{\XLinline}[1]{\textcolor{ForestGreen}{[XL: #1]}}
    \newcommand{\CT}[1]{\todo[fancyline,color=NavyBlue!40]{CT: #1}\xspace}
    \newcommand{\CTinline}[1]{\textcolor{NavyBlue}{[CT: #1]}}
    \newcommand{\AC}[1]{\todo[fancyline,color=Orchid!20]{AC: #1}\xspace}
    \newcommand{\ACinline}[1]{\textcolor{Orchid!20}{[AC: #1]}}
    
    \newcommand{\note}[1]{{\color{purple}[XL: #1]}}
    
    \newcommand{\pref}[1]{{\color{blue}(\ref{#1})}}
\fi

\newcommand{\Uinv}{U_{\mathrm{inv}}}        %
\newcommand{\Ufold}{U_{\mathrm{fold}}}      %

\newcommand{\Ephys}{E_{\mathrm{phys}}}

\newcommand{\wphys}{w_{\mathrm{phys}}}

\theoremstyle{plain}

\theoremstyle{definition}

\theoremstyle{remark}

%% file: abstract.tex
\begin{abstract}

Biomolecular design underpins applications from molecular recognition to therapeutics and synthetic biology, yet de novo interaction design remains challenging---especially for DNA/RNA, underexplored non-protein modalities with scarce, heterogeneous complex data and sharper geometric and chemical constraints.
We introduce MCTH (Monte Carlo Tree Hallucination), an inference-only framework that casts all-atom sequence--structure co-design as uncertainty-aware planning over hallucinated states from pretrained folding and inverse-folding models, with optional biophysical control within the same decision loop.
\algname treats these models as frozen black-box operators and uses Monte Carlo Tree Search to allocate a fixed inference budget across competing design trajectories, incorporating model confidence and uncertainty, as well as cross-expert consensus/disagreement when multiple predictors are available.
Across protein--RNA, protein--DNA, protein--protein, and protein--ligand design, matched-budget experiments show that adaptive search improves over simpler sampling and cycling strategies, while held-out AlphaFold3 and Chai-1 evaluations demonstrate transfer beyond the search-time oracle.
\algname provides a shared planning layer across modalities while allowing task-specific folding, inverse-folding, and biophysical modules, requiring no fine-tuning or backpropagation through component models.
\end{abstract}

%% file: introduction.tex
\section{Introduction}
\label{sec:intro}

Biomolecular design underpins modern molecular biology, synthetic biology, and biomedicine, enabling applications from catalysis and therapeutics to biosensing and programmable systems. A core goal is to design biomolecular interactions—protein–protein, protein–small-molecule, protein–nucleic-acid, and nucleic-acid–nucleic-acid complexes—with high structural fidelity and functional precision. Despite decades of progress, de novo design remains challenging, especially beyond proteins to DNA/RNA, small molecules, ions, and post-translational modifications.

A core difficulty is the intrinsically coupled nature of sequence and structure: in a rugged, high-dimensional landscape, small sequence changes can induce large structural or functional shifts. Capturing interaction specificity while jointly optimizing sequence and structure is computationally demanding and often difficult to control. {Physics-based modeling (e.g., Rosetta-style energy functions) provides explicit chemical priors but can be costly and brittle under local optimization {\citep{alford2017rosetta,leaverfay2011rosetta3}}. Learning-based methods offer expressive data-driven priors, yet often struggle with generalization, exploration, and optimization stability across heterogeneous modalities.} {These issues are amplified for DNA/RNA systems, where complex data are scarce and success hinges on fine-grained interaction fidelity (e.g., base stacking, ion/ligand coordination, atomic contacts) rather than global fold alone.}

Recent advances in deep learning–based structure prediction, culminating in AlphaFold3~\citep{abramson2024accurate}-style all-atom diffusion models, have expanded biomolecular modeling beyond proteins to small molecules, nucleic acids, ions, and covalent modifications. {Earlier predictors such as AlphaFold2 and AlphaFold-Multimer established strong protein(-complex) priors and intrinsic confidence measures {\citep{evans2021alphafoldmultimer, jumper2021highly}}. } However, structure prediction alone does not solve design. {Most design approaches either (i) use pretrained predictors as \emph{oracles} for validating or filtering large candidate pools~\citep{ferruz2022protgpt2,jendrusch2025alphadesign,liu2025scaffoldgpt,liu2025fragmentgpt, liu2025controllablegpt}, or (ii) turn predictors into optimization engines by \emph{fine-tuning}~\citep{liu2024scuba,watson2023rfdiffusion} or \emph{hallucination/inversion} objectives~\citep{anishchenko2021hallucination,cho2025boltzdesign1,pacesa2024bindcraft}.} Fine-tuned diffusion models require extensive training and are less flexible across tasks, while gradient-based hallucination approaches (e.g., bindcraft~\citep{pacesa2024bindcraft}, boltzdesign~\citep{cho2025boltzdesign1}) are expensive, initialization-sensitive, and prone to local optima—especially for long sequences and complex assemblies. {Backbone-generative diffusion models (e.g., RFdiffusion) offer geometric controllability but typically require separate sequence design and refolding-based verification \citep{dauparas2022proteinmpnn,watson2023rfdiffusion}.}

An emerging alternative paradigm exploits a property of diffusion-based structure prediction models often viewed as a limitation: hallucination. Given underspecified or out-of-distribution inputs—such as all-X sequences or partially defined complexes—these models can hallucinate plausible conformations guided by learned priors. Cycling between hallucinated structure prediction and sequence redesign can yield designable biomolecules without fine-tuning \fix{\citep{anishchenko2021hallucination,cho2025protein,fang2025halludesign}}. However, existing cycling-based methods rely on fixed schedules and greedy local updates, limiting global exploration and exploration–exploitation balance. {Moreover, high model confidence need not imply \emph{biophysical feasibility}: unconstrained hallucinations can violate fine-grained constraints (e.g., steric clashes, strained geometries, incompatible ligand/ion coordination).}

In this work, we propose \algname, an inference-time planning framework that reframes biomolecular design as structured search over hallucinated sequence–structure states. Pretrained folding and inverse-folding models act as frozen black-box operators that propose and evaluate candidate transitions, while Monte Carlo Tree Search allocates a fixed inference budget across competing design trajectories. Unlike fixed-depth cycling or gradient-based optimization through a structure predictor, MCTH requires only model outputs—sequences, structures, and confidence or feasibility scores—and does not access, update, or differentiate through component-model parameters.

The planner is uncertainty-aware: model confidence, expert consensus/disagreement when multiple predictors are available, and optional biophysical feasibility signals can jointly guide selection, expansion, and backup. The same planning logic is reused across modalities, while the folding, inverse-folding, and biophysical modules are dispatched according to the task. This yields a unified \emph{co-hallucination} strategy that couples learned generative proposals with biophysical control, which is particularly relevant for nucleic-acid and ligand-conditioned settings where data are limited and geometric constraints are sharp. In the main experiments we use a single all-atom folding model with modality-specific inverse folding; we additionally validate the multi-folding-expert capability in protein-binder design using Boltz-2 and AlphaFold2 jointly during search.

\paragraph{Contributions and novelty.}
\begin{itemize}
\setlength\itemsep{1pt}
\setlength\parskip{0pt}
\setlength\parsep{0pt}
\setlength\topsep{0pt}
\setlength\partopsep{0pt}
    \vspace{-0.4em}
    \item \textbf{Inference-time planning over sequence--structure states.} We cast hallucination-based biomolecular design as a finite-budget planning problem and use MCTS to adaptively allocate expensive folding evaluations across competing sequence--structure trajectories, rather than following fixed cycling or independent best-of-$N$ sampling.
    \item \textbf{Composable black-box model coordination.} MCTH treats pretrained folding, inverse-folding, and scoring models as frozen operators and requires only their emitted sequences, structures, and scalar signals---without fine-tuning or backpropagation through component models. Multiple folding experts can be combined through consensus, disagreement, and uncertainty signals.
    \item \textbf{Biophysics--ML co-hallucination:} We integrate optional geometric, thermodynamic, and energy-like feasibility cues with learned model confidence within the same planning loop, enabling task-specific control during selection, proposal refinement, and evaluation rather than only post-hoc filtering.
    \item \textbf{General planning layer across biomolecular design tasks.} The MCTS state, selection, expansion, and backup mechanism is shared across protein--RNA, protein--DNA, protein--protein, and protein--ligand design, while inverse-folding backends and biophysical terms remain modality-specific. Matched-budget and held-out cross-predictor evaluations test both search efficiency and robustness beyond the search-time oracle.
\end{itemize}

%% file: related_works.tex
\section{Background and Problem Setup}
\label{sec:background}

\paragraph{Related method families and positioning.}
Physics-first frameworks such as Rosetta/MD provide explicit energetic priors but scale poorly in large multimodal design spaces~\citep{alford2017rosetta,leaverfay2011rosetta3,lindorff2011fastfolding,rohl2004protein,shaw2010atomic}. Learning-based structure predictors (AF2/Multimer, all-atom AF3/RFAA) expanded biomolecular modeling beyond proteins~\citep{abramson2024accurate,evans2021alphafoldmultimer,jumper2021highly,krishna2024rfaa}; diffusion and inverse-folding tools such as RFdiffusion, ProteinMPNN, and RNA pipelines enabled generation/redesign~\citep{anand2025rnaframeflow,dauparas2022proteinmpnn,joshi2024grnade,joshi2025grnade,watson2023rfdiffusion}. Hallucination/cycling and planning methods, including BindCraft/BoltzDesign and ERP/MCTD-ME, improve optimization or exploration but are often local, fixed-schedule, or single-modality~\citep{anishchenko2021hallucination,cho2025boltzdesign1,cho2025protein,fang2025halludesign,Liu2024-ERP,liu2025monte,pacesa2024bindcraft}.

Relative to these lines, \algname occupies a distinct point in the design stack: it is a \emph{planning layer} that wraps pretrained folding and inverse-folding experts with plug-in biophysical cues and allocates expensive inference through MCTS. Its key distinction is \emph{co-hallucination}: learned experts propose sequence--structure hypotheses, while biophysical cues enter the same decision loop for selection, expansion, and proposal reweighting, rather than acting only as post-hoc filters. A fuller discussion of related work is provided in Appendix~\ref{app:related}, with a compact positioning summary in Appendix~\ref{app:positioning_summary}.

\vspace{-0.6em}
\paragraph{Sequence--structure design as planning.}
We represent a biomolecular system by sequence(s) $S$ and all-atom coordinates $X$; for complexes, the state includes all interacting components, e.g., $s=(S,X)$ under task conditioning $C$ such as binding partners, motifs, or ligands. We view design as finite-horizon planning in an MDP $\mathcal{M}=\langle \mathcal{S},\mathcal{A},H,P,R\rangle$~\citep{puterman2014markov}, where states $s\in\mathcal{S}$ are biomolecular sequence--structure hypotheses and actions $a\in\mathcal{A}$ are stochastic hallucination operators such as inverse-fold proposals, folding calls, or biophysical refinement/filtering steps. Transitions are induced by pretrained experts, $s'\sim P(\cdot\mid s,a)$, and rewards are computed from the resulting state, e.g., $R(s,a)=V(s')$.

\paragraph{Search and guided proposal view.}
MCTS iteratively expands a partial search tree through selection, expansion, evaluation, and backpropagation. In our setting, nodes are biomolecular states $s=(S,X)$, and evaluations come directly from model-derived value signals $V(s)$, such as confidence, interface quality, and optional feasibility terms, rather than long stochastic rollouts. Diffusion-style guidance can be viewed as reweighting proposals by an auxiliary score, $p^{\mathrm{guide}}(\xi'\mid\xi)\propto p(\xi'\mid\xi)\exp\{\beta f(\xi')\}$; in \algname, this score is instantiated by the same task-aware value signal used by planning. Practically, we implement this without gradients by sampling proposal candidates from different experts, seeds, or noise schedules and selecting/resampling according to their value.

\paragraph{Design objectives.}
We encode task objectives such as foldability, binding confidence, constraint satisfaction, and lead optimization through the state value \(V(s)\). Thus, MCTH searches over hallucinated sequence--structure states using learned confidence and optional biophysical feasibility signals, rather than performing explicit end-to-end energy minimization. Additional preliminaries are provided in Appendix~\ref{sec:app:supp_preliminaries}.

%% file: algorithm.tex
\section{Methodology}
\label{sec:method}

\begin{figure*}[t]
    \centering
    \includegraphics[width=\textwidth]{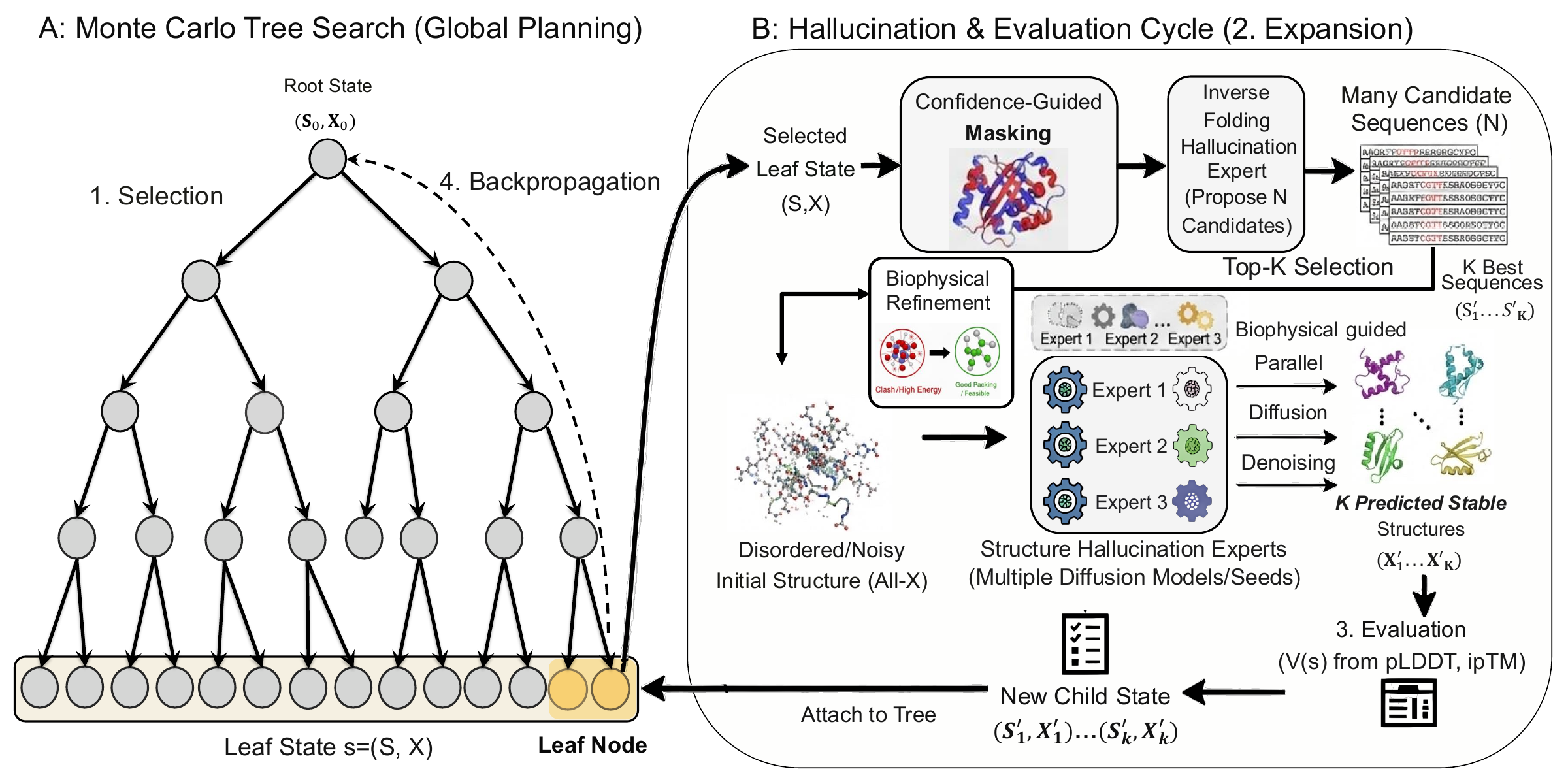}
    \caption{\fix{The \algname framework: search-guided sequence--structure co-design.}
    Left: UCT-style Monte Carlo Tree Search performs global planning over states $s=(S,X)$ via
    selection and backpropagation. Right: each expansion instantiates a hallucination--evaluation
    cycle that proposes new sequences (inverse folding), generates corresponding structures
    (diffusion-based hallucination), and evaluates candidates using model confidence signals (e.g.,
    ipTM/pLDDT), which are then backed up to guide subsequent search. \fix{Confidence-guided masking
    is an optional refinement operator and is disabled in the minimal plug-and-play configuration
    used in this work.}}
    \label{fig:mcth_flow}
    \vspace{-0.5cm}
\end{figure*}

We introduce \emph{Monte Carlo Tree Hallucination} (\algname), a unified framework for biomolecular
sequence--structure co-design that reframes hallucination by pretrained structure prediction models
as a principled \emph{search and planning problem}. Rather than relying on fixed iterative cycling
or gradient-based optimization, \algname formulates biomolecular design as structured exploration over
a joint sequence--structure landscape, guided by Monte Carlo Tree Search (MCTS) and driven by forward inference of pretrained folding and inverse-folding models together with biophysical guidance.

At a high level, \algname integrates three components: (i) diffusion-based all-atom structure hallucination, (ii) inverse-folding-based sequence redesign, and (iii) global exploration and decision-making via MCTS. \fix{\algname also incorporates plug-in biophysical feasibility/energy models (e.g., clash/packing/contact geometry penalties or energy-like terms) as complementary guidance signals, improving physics-aware controllability and physical fidelity without fine-tuning or backpropagation through structure predictors.}

\fix{Figure~\ref{fig:mcth_flow}} summarizes \algname as an \emph{uncertainty-aware} Monte Carlo Tree Search over
sequence--structure states: selection is guided by UCT-style visitation statistics and multi-expert
consensus/uncertainty signals, while each expansion instantiates a hallucination--evaluation cycle using
inverse folding and diffusion-based structure prediction, optionally interleaving biophysical scoring and refinement/filtering steps that enforce feasibility beyond model confidence.

\subsection{Composable Folding and Inverse-Folding Operators}
\label{sec:uncertainty_mcts}

\algname treats pretrained folding and inverse-folding models as frozen black-box operators that induce transitions in the search tree under task conditioning $C$ (e.g., binding partners, motifs, or ligands). The planner interacts with these models only through forward-inference outputs---predicted structures, proposed sequences, and associated confidence or feasibility signals---without accessing model parameters or backpropagating through the component models. This interface allows the same search procedure to compose different folding and inverse-folding backends while keeping the planner itself unchanged.

\paragraph{Structure hallucination via diffusion models.}
Given a sequence hypothesis $S$ and optional conditioning information $C$, a folding or structure-prediction model generates a candidate all-atom conformation
\begin{equation}
X \sim p_{\theta}(X \mid S, C).
\end{equation}
When the sequence or complex specification is underspecified---for example, an all-X sequence or a partially defined interface---the predictor may generate plausible conformations from its learned structural prior. \algname uses these predictions as candidate sequence--structure states for subsequent evaluation and search.

The framework supports either a single folding expert or multiple folding experts. In the main modality-spanning experiments, we use Boltz-2 as the shared all-atom folding expert together with modality-specific inverse-folding models. To directly validate the multi-expert extension, we additionally instantiate protein-binder search with both Boltz-2 and AlphaFold2. When multiple folding experts are instantiated, as in our dual-expert protein-binder experiment, we obtain an ensemble ${X'^{(e)}}_{e=1}^{E}$ for the same candidate sequence. Agreement among these predictions provides a consensus signal, while disagreement provides a structural-uncertainty signal. With a single folding expert, the same planner reduces naturally to confidence-based search without the cross-expert terms.

\paragraph{Sequence redesign via inverse folding.}
Given a hallucinated structure $X$ and conditioning information $C$ \fix{(e.g., binding partners, fixed motifs, ligand identity/pose constraints, or other task-specific constraints)}, inverse-folding experts propose compatible sequences
\begin{equation}
S \sim q_{\phi}(S \mid X, C),
\end{equation}
improving foldability and interaction consistency while preserving the specified constraints. Unlike fixed iterative pipelines, inverse folding in \algname is applied adaptively, allowing the planner to refine promising structural hypotheses or repair inconsistencies introduced by hallucination.

\subsection{Uncertainty-Aware MCTS over Sequence--Structure States}
\label{sec:uncertainty_mcts}

\algname casts biomolecular co-design as finite-budget tree search over
\emph{sequence--structure states}. Each node represents a concrete design hypothesis
\(s=(S,X)\), where \(S\) denotes the current sequence(s) of the designable component(s)
and \(X\) denotes the corresponding all-atom complex structure under task conditioning
\(C\) (e.g., fixed receptor, motif, ligand, or binding partner). This differs from
sequence-only search: a branch is evaluated not only by the proposed sequence, but by
the folded complex that the sequence induces. Thus, each edge corresponds to a
stochastic hallucination transition that composes sequence redesign and structure
prediction.

\paragraph{Actions and transitions.}
From a state \(s=(S,X)\), an action \(a\in\mathcal{A}(s)\) proposes a successor in two
steps. First, an inverse-folding expert proposes an updated design sequence \(S'\)
for the designable component, optionally with lightweight mutations or constraints.
Second, a folding/hallucination expert predicts the corresponding complex structure
\(X'\) under the same task conditioning \(C\), yielding a child state
\(s'=(S',X')\). When multiple folding experts, seeds, or noise schedules are used,
we obtain an ensemble \(\{X'^{(e)}\}_{e=1}^E\); agreement among these predictions
provides a consensus signal, while disagreement provides an uncertainty signal.

\paragraph{State value.}Each expanded child is assigned a task-aware value\begin{equation}V(s') = f!\left(m_1(s'),\ldots,m_J(s')\right),\end{equation}where the $m_j$'s are predictor-derived signals such as pLDDT, pTM, ipTM, interface quality, and optional geometric or biophysical feasibility terms. The aggregation function $f$ maps these signals to a scalar value used for evaluation and backup. Unless otherwise stated, we use a fixed task-level aggregation across targets rather than target-specific tuning. All signals are obtained through forward inference; \algname does not require gradients or backpropagation through the component models.

\paragraph{Uncertainty-aware selection.}
At each MCTS iteration, \algname traverses the current tree from the root to a leaf.
We use a \emph{consensus--uncertainty--physics} PUCT score, CU-PUCT:
\begin{align}
\mathrm{CU\mbox{-}PUCT}(s,a)
&=
Q(s,a)
+
c_p
\frac{\sqrt{\log N(s)}}{1+N(s,a)}
\cdot
\pi_{\mathrm{cons},\tau}(a\mid s)
\nonumber\\
&\quad \cdot
\Big(
w_{\mathrm{inv}}U_{\mathrm{inv}}(s,a)
+
w_{\mathrm{fold}}U_{\mathrm{fold}}(s,a)
\Big)
\cdot
\exp\!\left[-w_{\mathrm{phys}}\Ephys(s,a)\right].
\label{eq:cupuct_main}
\end{align}

We select
\begin{equation}
a^\star
=
\arg\max_{a\in\mathcal{A}(s)}
\mathrm{CU\mbox{-}PUCT}(s,a),
\end{equation}
where $Q(s,a)$ is the empirical action value, $N(s)$ and $N(s,a)$ are visit counts, and $c_p$ controls exploration.

When multiple folding experts are available, we define a consensus prior from their agreement on the same candidate transition. For a candidate sequence $S'$ and predicted structures $\{X'^{(e)}\}_{e=1}^{E}$,
\begin{equation}
\tilde{\pi}_{\mathrm{cons}}(s,a)
=
\exp\!\left(
\frac{1}{E}
\sum_{e=1}^{E}
\log q_{\phi}(S'\mid X'^{(e)},C)
\right)
\exp\!\left(
-\kappa
\mathcal{D}\!\left(\{X'^{(e)}\}_{e=1}^{E}\right)
\right),
\end{equation}
and $\pi_{\mathrm{cons},\tau}$ is obtained by temperature-normalizing
$\tilde{\pi}_{\mathrm{cons}}$ over candidate actions.
Thus, actions are preferred when the proposed sequence is compatible across predicted structures and the folding experts agree geometrically.

The uncertainty terms separate sequence-design ambiguity from structural uncertainty:
\begin{align}
U_{\mathrm{inv}}(s,a)
&=
\frac{1}{|M|}
\sum_{i\in M}
\mathsf{H}\!\left(
q_{\phi}(\cdot\mid X,C)_i
\right),
\\
U_{\mathrm{fold}}(s,a)
&=
g(\mathrm{pLDDT},\mathrm{pTM},\mathrm{ipTM},\ldots)
\quad\text{or}\quad
\mathcal{D}\!\left(\{X'^{(e)}\}_{e=1}^{E}\right).
\end{align}

Here, $U_{\mathrm{inv}}$ captures ambiguity in sequence redesign, while $U_{\mathrm{fold}}$ captures low folding confidence or, when multiple folding experts are used, cross-expert structural disagreement in the predicted complex. The optional biophysical term $\Ephys$ down-weights high-confidence but physically implausible proposals. Full normalization details and backup rules are given in Appendix~\ref{app:cupuct_details}.

\paragraph{Expansion, evaluation, and backup.}
Once selection reaches a leaf, \algname expands up to $K$ candidate children. Each child is constructed by inverse folding followed by folding/refolding, yielding a new sequence--structure state and the signals used to evaluate $V(s')$. The resulting value is backed up along the selected path to update visit counts and action-value statistics. Repeated iterations therefore allocate the fixed inference budget toward promising and informative branches and, when applicable, toward transitions supported across experts and consistent with biophysical constraints.

\paragraph{Remark 1: DNA aptamer design (thermodynamic window).}
For DNA aptamers, we instantiate \(\Ephys\) using a soft window penalty on
nearest-neighbor self-folding free energy \(\Delta G_{\mathrm{self}}(S')\), computed
at \(37\,^{\circ}\mathrm{C}\):
\begin{equation}
\begin{aligned}
\Ephys(S')
=
1 -
& \sigma\!\bigl(k_l(\Delta G_{\mathrm{self}}(S')-\Delta G_l)\bigr)
\sigma\!\bigl(-k_u(\Delta G_{\mathrm{self}}(S')-\Delta G_u)\bigr),
\end{aligned}
\label{eq:ephys}
\end{equation}
where \(\sigma(x)=1/(1+e^{-x})\) is the logistic sigmoid. We calibrate
\((\Delta G_l,\Delta G_u,k_l,k_u)\) from an empirical distribution over random
sequences; details and derivations are given in Appendix~\ref{app:param_derivation}.

\subsection{Biophysical Refinement}
\label{sec:biophys_refinement}

Biophysical signals in \algname can influence search not only through the state value
$\Ephys$ in Sec.~\ref{sec:uncertainty_mcts}, but also through optional refinement operators
that modify candidate states before subsequent refolding.
These refinements are task-specific and may act on sequence, structure, or both, while
leaving the underlying MCTS procedure unchanged.
For DNA aptamer design, we instantiate sequence-level refinement using nearest-neighbour
self-folding thermodynamics:

\paragraph{Identifying the failure mode.}
For each candidate $(S',X')$ output by MCTS, we compute $\Delta G_{\mathrm{self}}(S')$ using \texttt{seqfold}
and classify it into three regimes:
\begin{itemize}[noitemsep, wide,labelwidth=0pt, labelindent=0pt]
\setlength\itemsep{1pt}
\setlength\parskip{0pt}
\setlength\parsep{0pt}
\setlength\topsep{0pt}
\setlength\partopsep{0pt}
  \item \textbf{Over-stable} ($\Delta G_{\mathrm{self}} < \Delta G_l$): strong self-structure implies an
  unfolding penalty before binding.
  \item \textbf{Favourable} ($\Delta G_l \le \Delta G_{\mathrm{self}} \le \Delta G_u$): no refinement needed.
  \item \textbf{Under-structured} ($\Delta G_{\mathrm{self}} > \Delta G_u$): weak pre-organisation implies a
  high entropic cost upon binding.
\end{itemize}
Only over-stable or under-structured candidates are refined.
The boundaries $(\Delta G_l,\Delta G_u)$ are calibrated from an empirical distribution of random DNA sequences
(Appendix~\ref{app:param_derivation}).

\paragraph{Interface-aware mutation strategy.}
To avoid disrupting the binding surface, mutations are restricted to \emph{non-interface} positions.
We define an interface set $I \subset \{1,\ldots,|S'|\}$ as positions whose predicted contacts with the target in $X'$
exceed a threshold (e.g., heavy-atom distance $<8\,\text{\AA}$), and restrict edits to $j\notin I$.

{\emph{Over-stable correction.}}
We compute an MFE base-pair assignment $P(S')$ from \texttt{seqfold}, select a non-interface position $j^*$ with high
base-pairing propensity, and apply a point mutation $S'_{j^*}\!\to\!\tilde{s}$ chosen to locally destabilize the stem
(e.g., introduce a mismatch/bulge). We iterate until $\Delta G_{\mathrm{self}} \ge \Delta G_l$ or a budget of
$T_{\max}$ mutations is reached.

{\emph{Under-structured correction.}}
We identify a contiguous non-interface window (default length $w=6$) with low base-pairing propensity and propose a
stabilizing substitution that forms a short hairpin (e.g., a GC-rich stem with a stable tetraloop motif~\citep{santalucia1998unified}),
accepting the edit if it achieves $\Delta G_{\mathrm{self}} \le \Delta G_u$.

\paragraph{Acceptance criterion.}
We accept a refined sequence $S''$ iff it reduces the biophysical penalty:
\begin{equation}
\Ephys(S'') < \Ephys(S').
\label{eq:accept}
\end{equation}
Otherwise, we retain the original candidate $(S',X')$.

\subsection{Inference-time Structural Guidance}

Beyond value shaping and sequence refinement, biophysical information can optionally
enter a compatible folding model directly during inference.
This provides a third control interface in \algname: rather than only scoring or editing
a candidate, structural priors can steer the generation of its corresponding conformation.
This mechanism is backend-dependent and is not required by the MCTS planner.

For a candidate aptamer sequence $S'$, we run \texttt{seqfold} to obtain $\Delta G_{\mathrm{self}}(S')$ and an MFE base-pair set
$P(S')$; when $\Delta G_l \le \Delta G_{\mathrm{self}}(S') \le \Delta G_u$ (boundaries in Appendix~\ref{app:param_derivation}),
we enable Boltz-2 inference-time potentials (\texttt{use\_potentials}=\texttt{True}) and use $P(S')$ as a soft contact prior.
Specifically, we first identify the interface set $I$ from the current hallucinated complex $X'$ (positions contacting the
protein by a distance cutoff). We then retain only non-interface base pairs
$\mathcal{P}_{\text{free}}=\{(i,j)\in P(S'):\; i\notin I,\; j\notin I\}$ and impose the quadratic contact potential
\begin{equation}
U_{\text{contact}}(X)
\;=\;
\sum_{(i,j)\in\mathcal{P}_{\text{free}}}
w_{ij}\,\bigl(d_{ij}(X)-d_0\bigr)^2,
\label{eq:contact_potential}
\end{equation}
where $d_{ij}(X)$ is the heavy-atom distance for $(i,j)$ in structure $X$, $d_0$ is a target base-pair distance, and $w_{ij}$
can decay with diffusion noise level. Boltz-2 incorporates $U_{\text{contact}}$ via its potential-energy steering (e.g.,
$\hat{x}_0$ correction / particle resampling~\citep{del2006sequential}), requiring no backpropagation through the diffusion model.
This preserves secondary structure away from the binding site while leaving interface bases unconstrained for target binding. {Full implementation details (interface identification, contact scheduling, and potential weights) are provided in Appendix~\ref{sec:app:struct_guidance}.}

\subsection{Multi-modality Representation and Expert Assignment}
\label{sec:multimodal_main}

\algname separates the planning procedure from the models that instantiate its transition
and evaluation operators. The MCTS state representation, selection, expansion, and backup
rules are shared across tasks, while folding, inverse-folding, and optional biophysical
modules are assigned according to the molecular modality and available model interfaces.

For the modality-spanning main experiments, we use Boltz-2 as a shared all-atom folding
expert, whose representation supports proteins, nucleic acids, small molecules, and their
complexes~\citep{passaro2025boltz2}. Sequence redesign is dispatched by modality:
ProteinMPNN for proteins and peptides~\citep{dauparas2022proteinmpnn}, NA-MPNN for
nucleic acids~\citep{kubaney2025rna}, and LigandMPNN for ligand-conditioned protein
design. Task-specific biophysical terms and refinement operators are enabled only where
applicable.

The planner can also compose multiple compatible folding experts. In this setting,
candidate states may be evaluated by multiple predictors, allowing cross-expert consensus,
disagreement, and uncertainty to enter the same CU-PUCT selection rule defined above.
We evaluate this multi-expert instantiation separately in protein-binder design, while
keeping the underlying MCTS procedure unchanged.

Implementation details, supported modalities, and dispatch rules are summarized in
Appendix~\ref{sec:app:multimodal}.

%% file: experiments.tex
\section{Experiments}
\label{sec:experiments}
\paragraph{Evaluation protocol.}
We evaluate \algname from three complementary perspectives. First, we measure optimization under the \emph{search-time oracle}: candidates are evaluated using the same folding-model scores that define the inference-time design objective. Because \algname explicitly optimizes these scores, these experiments measure finite-budget optimization efficiency rather than oracle-independent physical validity.

Second, we use \emph{matched-budget controls} to separate adaptive search from raw evaluator access. Whenever applicable, baselines are given comparable candidate or folding-call budgets and evaluated under the same refolding protocol. We further compare against greedy cycling and best-of-$N$ variants using the same initialization, folding oracle, inverse-folding backend, and evaluation budget, isolating the effect of search-guided compute allocation.

Third, we perform \emph{held-out evaluation} using predictors and physical metrics that are not used to guide the corresponding search. We use AlphaFold3 for independent protein--nucleic-acid re-scoring, Chai-1 for held-out evaluation of the dual-expert protein-binder experiment, and Rosetta InterfaceAnalyzer as an additional oracle-independent physical check. For protein binders, we also report matched final-design pass rates under a held-out binder-quality filter rather than relying only on the single best-scoring candidate.

This distinction is important because some coupling to the search-time oracle is inherent to inference-time optimization: the planner must optimize a measurable objective. The relevant questions are therefore (i) whether adaptive search improves over equal-budget alternatives under the same objective, and (ii) whether the resulting designs remain competitive under evaluators outside the optimization loop.

\subsection{Nucleic Acid Design}
\label{sec:exp_na}

\subsubsection{RNA aptamer design}
\paragraph{Protein--RNA aptamer benchmark.}
We evaluate \algname on a protein--RNA aptamer design benchmark consisting of 7 protein--RNA chain-pair targets derived from PDB complexes
(7WKP, 7YEW, 7YGL, 8TG4, 8gxb, 9C7A), spanning RNA lengths from 4 to 61 nucleotides. Given a protein structure and the RNA length
specification, the goal is to design an RNA aptamer sequence that yields a confident protein--RNA interface.

\vspace{-0.6em}
\paragraph{Baselines and matched-budget evaluation.}
We compare against ODesign and RNAFrameFlow + NA-MPNN. To control for the effect of repeated structure evaluation, both baselines are run with 100 candidate designs and re-scored using the same Boltz-2 refolding protocol. All methods therefore receive a comparable opportunity to identify a high-scoring candidate under the search-time oracle.

\paragraph{Results under the search-time oracle.}
At the matched budget, \algname achieves a mean best ipTM of $0.946$, compared with $0.894$ for ODesign and $0.923$ for RNAFrameFlow + NA-MPNN. \algname obtains the highest ipTM on 6 of 7 targets; the only exception is 8GXB, where RNAFrameFlow is higher by $0.001$ ($0.880$ vs.\ $0.879$). Thus, increasing the baseline candidate pool substantially narrows the original gap but does not eliminate the advantage of adaptive search.

\paragraph{Held-out AlphaFold3 evaluation.}
We additionally re-score the final designs using AlphaFold3, which is never used during search. Under this held-out predictor, \algname retains the highest mean ipTM ($0.884$), compared with $0.824$ for ODesign and $0.794$ for RNAFrameFlow + NA-MPNN, and is top-ranked on 6 of 7 targets (including one tie). The persistence of the ranking under an independent predictor suggests that the gain is not confined to Boltz-2-specific preferences, although all results remain computational proxies rather than experimental validation.
\begin{table*}[t]
\centering
\small
\setlength{\tabcolsep}{5pt}
\renewcommand{\arraystretch}{1.05}

\begin{subtable}[t]{0.48\textwidth}
\centering
\begin{adjustbox}{max width=\linewidth}
\begin{tabular}{l r ccc}
\toprule
Target & RNA Len & ODesign@100 & RNAFrameFlow@100 & \algname \\
\midrule
7WKP\_A$\rightarrow$B & 22 & 0.883 & 0.932 & \textbf{0.941} \\
7YEW\_A$\rightarrow$B & 5  & 0.928 & 0.935 & \textbf{0.960} \\
7YGL\_A$\rightarrow$B & 4  & 0.918 & 0.924 & \textbf{0.963} \\
7YGL\_A$\rightarrow$C & 4  & 0.928 & 0.927 & \textbf{0.953} \\
8TG4\_A$\rightarrow$C & 5  & 0.924 & 0.925 & \textbf{0.963} \\
8GXB\_C$\rightarrow$B & 61 & 0.835 & \textbf{0.880} & 0.879 \\
9C7A\_A$\rightarrow$B & 8  & 0.841 & 0.940 & \textbf{0.960} \\
\midrule
\textbf{Mean} & -- & 0.894 & 0.923 & \textbf{0.946} \\
\bottomrule
\end{tabular}
\end{adjustbox}
\caption{Matched-budget Boltz-2 evaluation.}
\label{tab:rna_matched_budget}
\end{subtable}
\hfill
\begin{subtable}[t]{0.48\textwidth}
\centering
\begin{adjustbox}{max width=\linewidth}
\begin{tabular}{l ccc}
\toprule
Target & ODesign & RNAFrameFlow & \algname \\
\midrule
7WKP\_A$\rightarrow$B & 0.800 & 0.730 & \textbf{0.940} \\
7YEW\_A$\rightarrow$B & 0.880 & 0.890 & \textbf{0.900} \\
7YGL\_A$\rightarrow$B & 0.780 & 0.780 & \textbf{0.880} \\
7YGL\_A$\rightarrow$C & 0.780 & 0.800 & \textbf{0.870} \\
8TG4\_A$\rightarrow$C & \textbf{0.930} & 0.870 & \textbf{0.930} \\
8GXB\_C$\rightarrow$B & 0.770 & 0.680 & \textbf{0.850} \\
9C7A\_A$\rightarrow$B & \textbf{0.830} & 0.810 & 0.820 \\
\midrule
\textbf{Mean} & 0.824 & 0.794 & \textbf{0.884} \\
\bottomrule
\end{tabular}
\end{adjustbox}
\caption{Held-out AlphaFold3 evaluation.}
\label{tab:rna_af3}
\end{subtable}

\caption{
Protein--RNA aptamer design.
\textbf{Left:} matched 100-candidate evaluation under the Boltz-2 search-time oracle.
\textbf{Right:} independent AlphaFold3 re-scoring of final designs.
Values are ipTM; higher is better.
}
\label{tab:rna_aptamer_new}
\end{table*}

\subsubsection{DNA aptamer design}
\paragraph{Protein--DNA aptamer benchmark.}
We evaluate \algname on 10 protein--ssDNA chain-pair targets derived from four PDB complexes (7XVN, 7YSF, 7YUK, and 8PMF). Given a fixed protein target and a DNA length specification, the goal is to design an ssDNA aptamer sequence that forms a confident protein--DNA interface.

\paragraph{Baselines and matched-budget evaluation.}
We compare against ODesign as the native nucleic-acid design baseline. To control for repeated evaluator access, ODesign is run with 100 candidate designs and re-scored using the same Boltz-2 refolding protocol. Two 7XVN chain-B cases collapse to ipTM $\approx 0$ for both ODesign and \algname under Boltz-2; we therefore exclude these two oracle-failed cases from the matched-budget comparison and report the remaining eight chain-A targets. BoltzDesign is not included in this matched-budget table because its released design procedure optimizes the protein chain in protein--DNA complexes rather than directly generating the DNA strand.

\paragraph{Results under the search-time oracle.}
On the eight non-degenerate Boltz-2 targets, \algname achieves a mean best ipTM of $0.925$ versus $0.908$ for ODesign@100 and obtains the highest score on 7 of 8 targets. The gap is relatively small on several already high-scoring targets, indicating that strong best-of-$N$ sampling can approach search when the oracle landscape saturates, while adaptive allocation still improves the overall fixed-budget result.

\paragraph{Held-out AlphaFold3 evaluation.}
We additionally re-score designs from all 10 targets using AlphaFold3, which is never used during search. Under AF3, the two 7XVN chain-B cases are no longer degenerate, indicating that their collapse under Boltz-2 is evaluator-specific rather than evidence that the design targets themselves are intrinsically invalid. Across all 10 targets, \algname achieves the highest mean ipTM ($0.721$), compared with $0.665$ for BoltzDesign and $0.644$ for ODesign, and obtains the top score on 6 of 10 targets. Together with the matched-budget results, this shows that the advantage under the search-time oracle is not restricted to Boltz-2-specific scoring preferences, while also highlighting meaningful disagreement between structure predictors at the interface level.

\begin{table*}[t]
\centering
\small
\setlength{\tabcolsep}{5pt}
\renewcommand{\arraystretch}{1.05}

\begin{subtable}[t]{0.46\textwidth}
\centering
\begin{adjustbox}{max width=\linewidth}
\begin{tabular}{l cc}
\toprule
Target & ODesign@100 & \algname \\
\midrule
7XVN\_A$\rightarrow$N & 0.936 & \textbf{0.952} \\
7XVN\_A$\rightarrow$P & \textbf{0.957} & 0.950 \\
7YSF\_A$\rightarrow$B & 0.890 & \textbf{0.898} \\
7YSF\_A$\rightarrow$C & 0.869 & \textbf{0.876} \\
7YUK\_A$\rightarrow$C & 0.904 & \textbf{0.906} \\
7YUK\_A$\rightarrow$D & 0.901 & \textbf{0.912} \\
8PMF\_A$\rightarrow$B & 0.914 & \textbf{0.956} \\
8PMF\_A$\rightarrow$D & 0.895 & \textbf{0.947} \\
\midrule
\textbf{Mean} & 0.908 & \textbf{0.925} \\
\bottomrule
\end{tabular}
\end{adjustbox}
\caption{Matched-budget Boltz-2 evaluation on the 8 non-degenerate chain-A targets.}
\label{tab:dna_matched_budget}
\end{subtable}
\hfill
\begin{subtable}[t]{0.51\textwidth}
\centering
\begin{adjustbox}{max width=\linewidth}
\begin{tabular}{l ccc}
\toprule
Target & BoltzDesign & ODesign & \algname \\
\midrule
7XVN\_A$\rightarrow$N & 0.770 & \textbf{0.780} & 0.570 \\
7XVN\_A$\rightarrow$P & 0.740 & 0.570 & \textbf{0.850} \\
7XVN\_B$\rightarrow$N & \textbf{0.810} & 0.560 & 0.680 \\
7XVN\_B$\rightarrow$P & 0.440 & \textbf{0.740} & 0.670 \\
7YSF\_A$\rightarrow$B & 0.700 & 0.630 & \textbf{0.750} \\
7YSF\_A$\rightarrow$C & 0.410 & 0.470 & \textbf{0.680} \\
7YUK\_A$\rightarrow$C & \textbf{0.810} & 0.780 & 0.630 \\
7YUK\_A$\rightarrow$D & 0.630 & 0.780 & \textbf{0.820} \\
8PMF\_A$\rightarrow$B & 0.700 & 0.570 & \textbf{0.800} \\
8PMF\_A$\rightarrow$D & 0.640 & 0.560 & \textbf{0.760} \\
\midrule
\textbf{Mean} & 0.665 & 0.644 & \textbf{0.721} \\
\bottomrule
\end{tabular}
\end{adjustbox}
\caption{Held-out AlphaFold3 evaluation on all 10 targets.}
\label{tab:dna_af3}
\end{subtable}

\caption{
Protein--DNA aptamer design.
\textbf{Left:} matched 100-candidate evaluation under the Boltz-2 search-time oracle; the two 7XVN chain-B cases are excluded because Boltz-2 collapses for all compared designs.
\textbf{Right:} independent AlphaFold3 re-scoring on all 10 targets, under which the two chain-B cases are non-degenerate.
Values are ipTM; higher is better.
}
\label{tab:dna_aptamer_new}
\end{table*}

\subsection{Protein Binder Design}
\label{sec:exp_ppi}

We evaluate \algname on \emph{de novo} protein--protein binder design, aiming to generate a binder that forms a confident complex with a fixed receptor.

\paragraph{Benchmark and setup.}
We evaluate \algname on de novo protein--protein binder design using the 12-target CaoData benchmark. For each target, the receptor is fixed with its native sequence and structure, while the binder is designed de novo at the reference length. Targets span diverse receptor and binder sizes, including the multi-chain antibody-style case 5U8R.

For the main matched-budget experiment, we instantiate \algname with Boltz-2 as the folding/refolding expert and ProteinMPNN as the inverse-folding proposal model. Starting from a random binder sequence at the target length, \algname searches over sequence--structure hypotheses by alternating structure prediction and structure-conditioned sequence redesign. Unless otherwise stated, the search budget is 100 Boltz-2 evaluations per target.

\paragraph{Matched-budget optimization.}
We first test whether the advantage of \algname can be explained by greater access to the folding evaluator. We therefore compare against BoltzDesign1, ODesign, and Complexa under a matched budget of up to 100 folding/evaluator calls per target. ODesign and Complexa candidates are compared under the common Boltz-2 refolding protocol, while BoltzDesign1 uses its native Boltz evaluation during optimization.

Across CaoData, \algname achieves the highest mean best ipTM ($0.864$), compared with $0.853$ for Complexa, $0.816$ for ODesign, and $0.539$ for BoltzDesign1 (11 available targets). Thus, the gain cannot be attributed simply to receiving more folding-model evaluations.

\begin{table}[t]
\centering
\small
\setlength{\tabcolsep}{4pt}
\begin{tabular}{lcccc}
\toprule
Target & BoltzDesign1@100 & ODesign@100 & Complexa@100 & \algname@100 \\
\midrule
1DJS & 0.758 & \textbf{0.944} & 0.907 & 0.933 \\
1MOX & 0.291 & 0.901 & 0.845 & \textbf{0.929} \\
1XIW & 0.884 & 0.940 & \textbf{0.946} & \textbf{0.946} \\
2GY7 & 0.189 & 0.672 & \textbf{0.930} & 0.862 \\
2IFG & 0.527 & 0.730 & 0.356 & \textbf{0.806} \\
3DI3 & 0.508 & 0.901 & 0.926 & \textbf{0.947} \\
3FKD & 0.588 & 0.556 & \textbf{0.958} & 0.686 \\
3MJG & 0.697 & 0.878 & 0.888 & \textbf{0.904} \\
3ZTJ & 0.568 & 0.872 & \textbf{0.941} & 0.914 \\
4O3V & 0.778 & 0.944 & 0.852 & \textbf{0.953} \\
4OGA & 0.136 & \textbf{0.916} & 0.904 & 0.912 \\
5U8R & -- & 0.532 & \textbf{0.783} & 0.576 \\
\midrule
\textbf{Mean} & 0.539$^\dagger$ & 0.816 & 0.853 & \textbf{0.864} \\
\bottomrule
\end{tabular}
\caption{
Matched-budget protein-binder optimization on CaoData.
All methods use up to 100 folding/evaluator calls per target.
ODesign and Complexa candidates are evaluated under the common Boltz-2 refolding protocol, while BoltzDesign1 uses its native Boltz evaluation during optimization.
$^\dagger$BoltzDesign1 mean is over 11 targets because 5U8R is unavailable.
Values are best ipTM per target; higher is better.
}
\label{tab:protein_matched_budget}
\end{table}

\paragraph{Dual-expert cross-predictor evaluation.}
We next test whether incorporating multiple folding experts improves robustness beyond a single search-time oracle. We retain the same ProteinMPNN proposal mechanism and MCTS procedure, but evaluate candidate states using both Boltz-2 and AlphaFold2. Their structural-quality signals contribute to node evaluation, while cross-expert consensus, disagreement, and uncertainty enter CU-PUCT exploration. Chai-1 is held out entirely from search and is used only for final evaluation.

Across the 11 targets shared by all configurations, the original Boltz-only \algname obtains mean ipTM values of $0.69/0.19/0.18$ under Boltz-2/AF2/Chai-1, respectively. The dual-expert variant improves these to $0.74/0.61/0.57$. In particular, held-out Chai-1 improves from $0.18$ to $0.57$, with the dual-expert variant outperforming the single-expert configuration on 9 of 11 targets and tying on one. Thus, incorporating a second structural expert substantially reduces predictor-specific optimization and improves transfer to a third predictor outside the search loop.

\begin{table*}[t]
\centering
\small
\setlength{\tabcolsep}{4pt}
\renewcommand{\arraystretch}{1.05}
\begin{adjustbox}{max width=\textwidth}
\begin{tabular}{l ccc ccc ccc}
\toprule
\multirow{2}{*}{Target}
& \multicolumn{3}{c}{Single-expert \algname}
& \multicolumn{3}{c}{Dual-expert \algname}
& \multicolumn{3}{c}{RFDiffusion} \\
\cmidrule(lr){2-4}
\cmidrule(lr){5-7}
\cmidrule(lr){8-10}
& B2 & AF2 & Chai
& B2 & AF2 & Chai
& B2 & AF2 & Chai \\
\midrule
1DJS & 0.89 & 0.19 & 0.16 & 0.86 & 0.79 & \textbf{0.73} & 0.39 & 0.28 & 0.59 \\
1MOX & 0.70 & 0.14 & 0.15 & 0.23 & 0.48 & 0.15 & 0.27 & 0.24 & \textbf{0.50} \\
1XIW & 0.60 & 0.39 & 0.34 & \textbf{0.95} & \textbf{0.86} & \textbf{0.84} & 0.77 & 0.84 & 0.81 \\
2GY7 & 0.31 & 0.21 & 0.17 & \textbf{0.90} & \textbf{0.68} & \textbf{0.76} & 0.37 & 0.67 & 0.73 \\
2IFG & 0.72 & 0.18 & 0.10 & \textbf{0.76} & 0.15 & 0.59 & 0.30 & \textbf{0.78} & \textbf{0.70} \\
3DI3 & \textbf{0.90} & 0.12 & \textbf{0.34} & 0.76 & \textbf{0.74} & 0.22 & 0.58 & 0.62 & 0.26 \\
3FKD & \textbf{0.52} & 0.23 & 0.14 & 0.43 & 0.17 & 0.19 & 0.32 & \textbf{0.88} & \textbf{0.31} \\
3MJG & 0.50 & 0.13 & 0.12 & \textbf{0.82} & \textbf{0.61} & \textbf{0.64} & 0.47 & 0.16 & 0.53 \\
3ZTJ & \textbf{0.70} & 0.19 & 0.13 & 0.67 & \textbf{0.80} & \textbf{0.74} & 0.35 & 0.56 & 0.66 \\
4O3V & 0.93 & 0.16 & 0.26 & \textbf{0.94} & \textbf{0.83} & \textbf{0.82} & 0.68 & 0.75 & 0.78 \\
4OGA & 0.77 & 0.11 & 0.09 & \textbf{0.87} & 0.62 & 0.63 & 0.71 & \textbf{0.66} & \textbf{0.69} \\
\midrule
\textbf{Mean}
& 0.69 & 0.19 & 0.18
& \textbf{0.74} & \textbf{0.61} & 0.57
& 0.47 & 0.58 & \textbf{0.60} \\
\bottomrule
\end{tabular}
\end{adjustbox}
\caption{
Cross-predictor evaluation on the 11 shared CaoData protein-binder targets.
Single-expert \algname uses Boltz-2 during search, whereas dual-expert \algname uses both Boltz-2 and AlphaFold2; Chai-1 is held out from search.
For \algname, each evaluator column reports the best ipTM over three retained designs; RFDiffusion reports the best over ten designs.
Higher is better.
}

\label{tab:dual_expert_cross_oracle}
\end{table*}

\paragraph{Matched final-design pass rate.}
Best-per-target scores characterize whether a method can discover at least one strong candidate, but do not measure design yield. We therefore additionally compare \algname and RFDiffusion using exactly 10 final designs per target on the same 11-target set and apply the same held-out binder-quality filter to both methods. \algname candidates are selected using only the recorded in-loop joint reward; held-out metrics are not used retrospectively for ranking.

Dual-expert \algname yields 41 passing designs out of 110 ($37.3\%$), compared with 22/110 ($20.0\%$) for RFDiffusion, corresponding to a $17.3$ percentage-point or $1.86\times$ higher pass rate. Both methods produce at least one passing design on 8 of 11 targets. Thus, while RFDiffusion can achieve a slightly higher best-of-output Chai score, \algname produces a substantially larger fraction of qualifying designs under an equal output budget.

\begin{table}[t]
\centering
\small
\begin{tabular}{lcccc}
\toprule
Method & Outputs/target & Passing/valid & Pass rate & Targets $\geq$1 pass \\
\midrule
Dual-expert \algname & 10 & 41/110 & \textbf{37.3\%} & 8/11 \\
RFDiffusion & 10 & 22/110 & 20.0\% & 8/11 \\
\bottomrule
\end{tabular}
\caption{Matched final-design yield under the same held-out binder-quality filter.}
\label{tab:protein_passrate}
\end{table}

%% file: conclusion.tex
\section{Conclusion}
\label{sec:conclusion}

We presented \algname, an inference-only framework that reframes biomolecular design as an \emph{uncertainty-aware} search over hallucinated sequence--structure states. By coordinating frozen folding, inverse-folding, and optional biophysical operators with MCTS, \algname allocates expensive structure evaluations using confidence, uncertainty, and, when multiple experts are available, consensus/disagreement signals. Across protein--RNA, protein--DNA, protein--protein, and protein--ligand design, matched-budget experiments demonstrate effective search-guided compute allocation, while held-out AlphaFold3 and Chai-1 evaluation shows that improvements can transfer beyond the search-time oracle. The planning layer is shared across tasks, while expert backends and biophysical terms remain modality-specific. Our evaluation remains entirely computational and does not replace experimental validation. Future work should prioritize wet-lab validation, stronger physics-based objectives, and broader multi-expert evaluation across modalities. Overall, \algname suggests that closed-loop, biophysics-informed planning is a promising route for controllable biomolecular co-design across heterogeneous modalities.

\clearpage

%% file: supp_related_works.tex
\section{Related works}
\label{app:related}

\paragraph{Physics-Based and Knowledge-Driven Biomolecular Design.}
Early approaches to biomolecular design were dominated by physics-based energy modeling and sampling, exemplified by frameworks such as Rosetta {and its modern all-atom energy functions ~\citep{rohl2004protein,leaverfay2011rosetta3,alford2017rosetta}.}  These methods provide explicit physical interpretability and fine-grained energetic control, but suffer from high computational cost and limited scalability when exploring large sequence spaces or multimolecular complexes. {Molecular dynamics (MD) provides an alternative physics-first paradigm for atomistic refinement and thermodynamic reasoning~\citep{shaw2010atomic,lindorff2011fastfolding},
yet remains computationally costly for design-time exploration at scale, especially for long sequences and heterogeneous complexes.} While hybrid approaches combining coarse-grained sampling with heuristic pruning have improved tractability, global exploration of the coupled sequence–structure landscape remains challenging.

\paragraph{Learning-Based Structure Prediction and Generative Design.}
Deep learning–based structure prediction~\citep{peng2011raptorx, lin2023evolutionary}  has fundamentally transformed biomolecular modeling. {AlphaFold2 and AlphaFold-Multimer established strong protein(-complex) priors~\citep{jumper2021highly,evans2021alphafoldmultimer} and RoseTTAFold provided a complementary open predictor~\citep{baek2021accurate}.} {More recently, generalized all-atom models extend prediction to assemblies containing proteins, nucleic acids, ligands/ions, and covalent modifications, including RoseTTAFold All-Atom~\citep{krishna2024rfaa} and AlphaFold3~\citep{abramson2024accurate}. {Concurrently, ODesign proposes an AlphaFold3-like all-atom ``world model'' for all-to-all biomolecular interaction design across proteins, nucleic acids, and small molecules, with multi-level controllability via masking and conditional generation~\citep{zhang2025odesign}.} These models provide powerful learned structural priors and intrinsic confidence signals, but are primarily trained for prediction rather than optimization; repurposing them for controllable design still requires effective exploration and a principled mechanism for incorporating external constraints.}

{\paragraph{Generative design with diffusion and inverse folding.}
Diffusion and flow-matching models enable direct backbone/structure generation, with RFdiffusion providing a general framework for protein generation and binder/motif-conditioned design~\citep{watson2023rfdiffusion}. Sequence design models such as ProteinMPNN provide efficient inverse folding for fixed backbones and are widely used as downstream redesign modules~\citep{dauparas2022proteinmpnn}. For nucleic acids, emerging geometric generators and inverse-design pipelines include gRNAde (3D RNA inverse design) and RNA-FrameFlow (RNA backbone generation) \citep{joshi2024grnade,joshi2025grnade,anand2025rnaframeflow}, highlighting both the promise and the current fragmentation of modality-specific tooling. For protein--ligand settings, diffusion has also been used to model ligand poses and docking distributions (e.g., DiffDock)~\citep{corso2023diffdock}. A common limitation across these approaches is that design quality often depends on large candidate pools plus refolding-based verification, and controllable global exploration remains challenging when objectives are multi-constraint or when the model is used out-of-distribution.}

\paragraph{Gradient-based and iterative hallucination methods.}
A complementary line of work turns pretrained predictors into optimizers via backpropagation through confidence surrogates or via iterative cycling between structure prediction and sequence redesign. BindCraft demonstrates strong empirical performance for protein binder design via AF2-based gradient optimization and filtering~\citep{pacesa2024bindcraft}, while BoltzDesign1 inverts an all atom predictor (Boltz-1, AF3-like) to optimize distograms for generalized binder design \citep{cho2025boltzdesign1}. Cycling-based ``hallucination'' pipelines that repeatedly refold and redesign can avoid fine-tuning and yield designable structures in practice~\citep{anishchenko2021hallucination,cho2025protein,fang2025halludesign}. However, these methods are typically local: they rely on greedy acceptance, fixed schedules, or initialization-sensitive continuous optimization, and can be computationally expensive or unstable for long sequences and heterogeneous complexes. Moreover, constraint handling is often implicit (through heuristics) rather than explicitly controllable at the algorithmic level.

\paragraph{Search and planning for molecular generation.}
Search-based methods are well studied in Reinforcement Learning (RL) and combinatorial optimization, but are less explored for continuous, multimodal biomolecular design due to large branching factors, delayed rewards, and expensive evaluations. ERP integrates MCTS with language-model generation for ligands, improving exploration over greedy decoding~\citep{Liu2024-ERP}, while MCTD-ME couples masked diffusion with MCTS for protein \emph{sequence} planning and long-range revision~\citep{liu2025monte}. More broadly, RL-style generative frameworks (e.g., GFlowNets) aim to sample diverse high-reward objects under complex objectives~\citep{bengio2021gflownet}. {Despite these advances, prior planning-based methods are typically limited to a single modality (often sequences or SMILES) and do not unify all-atom sequence--structure co-design across proteins, nucleic acids, and ligand-containing complexes.}

\paragraph{Positioning of \algname.}
Relative to (i) physics-first energy optimization pipelines (e.g., Rosetta/MD), (ii) direct all-atom generators (e.g., RFdiffusion and ODesign), and (iii) gradient-based or fixed-schedule hallucination/cycling methods (e.g., BindCraft, BoltzDesign), \algname occupies a distinct point in the design stack: it is a \emph{planning layer} that wraps pretrained folding and inverse-folding experts \fix{together with plug-in biophysical models (e.g., energy/geometry feasibility and clash/packing/contact penalties)} and allocates expensive inference via MCTS.
Our key conceptual distinction is \emph{co-hallucination}: learned experts provide expressive structural proposals, while plug-in biophysical feasibility/energy cues enter the \emph{same} decision loop (selection/expansion/proposal reweighting) as controllable guidance signals, rather than as purely post hoc filters.
This separation of concerns—experts for proposal, planning for compute allocation, and physics for controllable feasibility—enables a unified, fine-tuning-free workflow that remains applicable in data-sparse regimes such as nucleic-acid and ligand-conditioned design.

\subsection{Positioning Summary}
\label{app:positioning_summary}

Table~\ref{tab:positioning_summary} summarizes the high-level positioning of MCTH relative to representative method families. The goal of this comparison is not to claim that any individual ingredient is new in isolation, but to clarify the specific combination emphasized by MCTH: planning over explicit sequence--structure states, broad modality coverage, biophysical feasibility inside the decision loop, and no task-specific fine-tuning.

\begin{table}[h]
\centering
\caption{High-level comparison with related method families. A check mark indicates an explicit/core component, $\triangle$ indicates partial or task-specific support, and $\times$ indicates absence.}
\label{tab:positioning_summary}
\small
\begin{tabularx}{\linewidth}{L>{\centering\arraybackslash}p{0.18\linewidth}>{\centering\arraybackslash}p{0.14\linewidth}>{\centering\arraybackslash}p{0.15\linewidth}>{\centering\arraybackslash}p{0.14\linewidth}}
\toprule
Method family / example & Plans over seq.--struct. states & Multi-modality & Biophysics in-loop & Fine-tuning-free \\
\midrule
Rosetta / MD-style optimization & $\times$ & Limited & \checkmark & \checkmark \\
RFdiffusion / ODesign & $\times$ & $\triangle$ & $\times$ & \checkmark \\
ProteinMPNN / inverse folding & $\times$ & Limited & $\times$ & \checkmark \\
BindCraft / BoltzDesign & $\triangle$ & Limited & $\triangle$ & \checkmark \\
MCTD-ME & \checkmark & Protein only & $\times$ & \checkmark \\
\textbf{MCTH} & \checkmark & \checkmark & \checkmark & \checkmark \\
\bottomrule
\end{tabularx}
\end{table}

%% file: supp_algorithm.tex
\section{Additional Methodological Details}
\label{sec:app:method_details}

\subsection{Uncertainty-Aware MCTS over Sequence--Structure States}
\label{app:cupuct_details}

\fix{Following Sec.~\ref{sec:background}, we represent a biomolecular design hypothesis by a state $s=(S,X)$, where $S$ denotes the current (possibly underspecified) sequence(s) of all designable components and $X$ denotes the corresponding all-atom complex structure under task conditioning $C$.} MCTH casts design as a finite-budget planning problem in which the budget is dominated by expensive folding calls.

\paragraph{Actions and transitions.}
From state $s$, an action $a\in\mathcal{A}(s)$ proposes a successor by: (i) proposing an
updated design sequence $S'$ for the designable component (e.g., via inverse folding, optionally
with lightweight mutations), followed by (ii) invoking a folding/hallucination expert to obtain a
corresponding complex structure under task conditioning $C$ (fixed receptor, ligand specification,
motif constraints, etc.), yielding $s'=(S',X')$.
{When multiple folding/hallucination experts are available, we instantiate an ensemble
$\{X'^{(e)}\}_{e=1}^E$ and use cross-expert agreement as an explicit uncertainty signal.}

\paragraph{Node value (reward) from model confidence.}
Each expanded state is evaluated by a task-aware value function
\begin{equation}
V(s) \;=\; f\!\big(m_1(s),\ldots,m_J(s)\big),
\end{equation}
where $\{m_j\}$ are intrinsic signals produced by pretrained predictors, such as global fold
confidence (e.g., pLDDT/pTM), interface confidence (e.g., ipTM for binding tasks), and optional
geometric or contact-based terms when available. The aggregation function $f$ maps these raw
predictor outputs to a scalar state value used for MCTS selection, evaluation, and backup. In our
experiments, unless otherwise stated, $f$ is instantiated as a fixed linear aggregation of confidence
and interface-quality signals, with weights reported in Appendix~\ref{app:hyperparams}. The
aggregation is task-aware but not target-tuned: the same weights are used across benchmarks unless
a task-specific term, such as a motif constraint, is absent and therefore set to zero. Importantly,
$V(s)$ is computed using only forward inference outputs; no gradients, fine-tuning, or
backpropagation through structure predictors is used.

\paragraph{Search procedure (\fix{uncertainty-aware CUPUCT-MCTS}).}
MCTH performs Monte Carlo Tree Search rollouts under a fixed compute budget (e.g., a maximum number
of expensive folding calls). Each rollout consists of four stages: selection, expansion, evaluation,
and backpropagation.

\emph{Selection.}
Starting at the root, we recursively select child edges to traverse until reaching a leaf node.
\fix{We introduce a \emph{consensus--uncertainty--physics} PUCT-style score, \CUPUCT, where the exploration term is
modulated by (i) a \emph{consensus prior} $\pi_{\mathrm{cons},\tau}(\action\mid \state_t)$ and (ii) an
\emph{uncertainty--feasibility} factor from inverse/forward-fold experts, with an optional biophysical
penalty $E_{\mathrm{phys}}(\state_t,\action)$.}
\fix{We compute}
\[
\fix{
\begin{aligned}
\CUPUCT\!& \left((\state_t,\action)\right)
= \QFunc(\state_t,\action)
+ c_p\,\frac{\sqrt{\log \Num(\state_t)}}{1+\Num(\state_t,\action)} \cdot \pi_{\mathrm{cons},\tau}(\action\mid \state_t) \\
&\quad \cdot \Big( w_{\mathrm{inv}}\,\Uinv(\state_t,\action)
\;+\; w_{\mathrm{fold}}\,\Ufold(\state_t,\action) \Big) \cdot \exp\!\Big(\!-\;w_{\mathrm{phys}}\,E_{\mathrm{phys}}(\state_t,\action)\Big),
\end{aligned}
}
\]
and select $\action^*(\state_t)=\arg\max_{\action\in\mathcal{A}(\state_t)} \CUPUCT((\state_t,\action))$.
Here $\QFunc$ is the empirical action-value estimate, $\Num(\state_t)$ and $\Num(\state_t,\action)$ are node/edge visit counts,
and $c_p$ controls the exploration--exploitation trade-off.
\fix{The terms $\pi_{\mathrm{cons},\tau}$, $\Uinv$, $\Ufold$, and $E_{\mathrm{phys}}$ are computed once at expansion and cached,
keeping selection lightweight.}

{\emph{Consensus prior.}}
{We define $\pi_{\mathrm{cons},\tau}(\action\mid\state_t)$ to favor actions whose transition is
supported by multiple experts. For a candidate child produced by inverse folding $S'$ and an ensemble
$\{X'^{(e)}\}_{e=1}^E$, we form an unnormalized consensus score}
\fix{
\[
\tilde{\pi}_{\mathrm{cons}}(\state_t,\action)
=
\exp\!\Big(\frac{1}{E}\sum_{e=1}^E \log q_{\phi}(S' \mid X'^{(e)},C)\Big) \cdot
\exp\!\big(-\kappa\,\mathcal{D}(\{X'^{(e)}\})\big),
\]
}
{where $q_{\phi}$ is the inverse-folding likelihood and $\mathcal{D}(\{X'^{(e)}\})$ is a cross-expert
disagreement measure (e.g., backbone/interface dispersion). We then apply temperature-$\tau$ normalization}
\begin{equation}
{
\pi_{\mathrm{cons},\tau}(\action\mid\state_t)
=
\frac{\tilde{\pi}_{\mathrm{cons}}(\state_t,\action)^{1/\tau}}
{\sum_{\action'} \tilde{\pi}_{\mathrm{cons}}(\state_t,\action')^{1/\tau}}.
}
\end{equation}
{This yields an action prior that interpolates between uniform exploration (large $\tau$) and sharp
consensus-driven focus (small $\tau$).}

{\emph{Uncertainty terms.}}
{We decompose uncertainty into two sources. First, inverse-folding uncertainty}
\begin{equation}
{
\Uinv(\state_t,\action)
=
\frac{1}{|M|}
\sum_{i\in M}
\mathsf{H}\!\left(q_{\phi}(\cdot \mid X,C)_i\right),
}
\end{equation}
where $\mathsf{H}(\cdot)$ is categorical entropy and $M$ is the set of updated positions
\emph{when masking is disabled, $M$ is the full designable set}). Second, folding uncertainty is defined either from confidence-derived uncertainty or from cross-expert structural disagreement:
\begin{align}
\Ufold(\state_t,\action)= g\!\big(\text{pLDDT},\text{pTM},\text{ipTM},\ldots\big), \Ufold(\state_t,\action)=\mathcal{D}(\{X'^{(e)}\}).
\end{align}
Here, $g(\cdot)$ maps folding confidence outputs to an uncertainty bonus. In practice, $g$ is implemented as a monotone decreasing transformation of confidence, so that lower confidence or poorer interface resolution yields larger $\Ufold$ and encourages exploration of under-resolved states. The alternative form $\mathcal{D}(\{X'^{(e)}\})$ captures epistemic disagreement across folding experts, where larger dispersion among predicted structures indicates greater folding uncertainty.

Finally, we include an optional biophysical term $E_{\mathrm{phys}}$ to discourage nonphysical hypotheses. Overall, CUPUCT-MCTS provides an uncertainty-aware planning layer: $\pi_{\mathrm{cons},\tau}$ stabilizes search toward expert-agreed transitions, while $\Uinv$ and $\Ufold$ direct exploration toward uncertain or under-resolved regions of the coupled sequence--structure landscape.

{\emph{Biophysical feasibility term.}}
\fix{Finally, we incorporate an optional biophysical feasibility/energy factor $E_{\mathrm{phys}}(\state_t,\action)$
to down-weight proposals that are high-confidence under learned priors but physically implausible (e.g., clashes, strained
geometry, infeasible pockets). We define a modular biophysical score $\Ephys(\cdot)\in[0,1]$ (near $0$ if feasible; near $1$
if implausible) and set $E_{\mathrm{phys}}(\state_t,\action)\triangleq \Ephys(S',X')$ for the child produced by $\action$,
computed from the proposed sequence $S'$ and/or predicted coordinates $X'$ \emph{without additional folding calls}. This
term enters \CUPUCT~multiplicatively as $\exp(-w_{\mathrm{phys}}\,E_{\mathrm{phys}})$ and is cached at expansion time.}

\emph{Expansion.}
Upon reaching a leaf, we expand it by sampling up to $K$ candidate actions (e.g., $K$ sequence proposals)
and adding the resulting successor states as children. Each child is constructed by running the
inverse-folding expert (to propose $S'$) and then the folding/hallucination expert (to obtain $X'$),
producing a concrete complex hypothesis.
{At expansion time, we compute and cache $\pi_{\mathrm{cons},\tau}(\action\mid\state_t)$,
$\Uinv(\state_t,\action)$, $\Ufold(\state_t,\action)$, and (when available) $E_{\mathrm{phys}}(\state_t,\action)$,
which are then reused by the selection policy.}

\emph{Evaluation.}
Newly added children are scored by computing $V(s')$ from the folding expert’s confidence outputs
(and any task-specific geometric checks). In our implementation, this evaluation is performed
immediately after expansion (i.e., the expensive folding call doubles as the rollout evaluation),
keeping the planner modular and plug-and-play.

\emph{Backpropagation.}
The evaluated value is propagated back along the selected path to update visit counts and value
statistics. Concretely, we update $\Num(\cdot)$ and maintain $\QFunc(\cdot)$ via a standard backup rule
(e.g., running mean or max-backup), enabling the tree policy to increasingly favor high-value branches
over repeated simulations.

\subsection{Biophysical Penalty}
\label{app:biophys_reward}

The confidence signals $V(s)$ (such as pLDDT and ipTM) measure how
consistent a predicted structure is with the folding expert's learned priors,
but they do not capture all physical constraints relevant to real-world
binding or function.
Different biomolecular design tasks impose different task-specific physical
requirements---thermodynamic stability windows for nucleic acid aptamers,
steric feasibility for protein--ligand interfaces, geometric packing quality
for de novo protein design---that are invisible to a folding expert evaluating
only the bound complex.

In addition to the ML confidence signal, we introduce a modular
\emph{biophysical energy term} $\Ephys(S') \in [0,1]$ that encodes
task-specific physical feasibility.
$\Ephys$ is near zero for physically plausible candidates and approaches one
for implausible ones; it enters the PH-UCB selection rule as a multiplicative
factor $\exp(-\wphys\,\Ephys)$, acting as a soft preference ($\wphys = 0.1$)
rather than a hard filter.
The specific form of $\Ephys$ is defined per task and can be computed from
the sequence alone or from predicted coordinates, without requiring an
additional structure prediction call.

\paragraph{Example: DNA aptamer design.}
As a concrete example, consider DNA aptamer design.
Systematic thermodynamic measurements show that both extremes of self-folding
stability hurt binding: aptamers with very stable self-folds must unfold
before engaging the target, paying an unfolding cost that reduces effective
affinity, while aptamers with too little intrinsic structure lose the
pre-organisation benefit and pay a steep entropic cost upon
binding~\citep{alkhamis2025exploring, vallee2009thermodynamic}.
Neither failure mode is visible to the folding expert, which only evaluates
the bound complex.

For a candidate aptamer sequence $S'$, we define $\Ephys$ as
a soft window penalty on the nearest-neighbour minimum free energy:
\begin{equation}\label{eq:ephys}
\begin{aligned}
\Ephys(S')
\;=\;
1 \;-\;
&\sigma\!\bigl(k_l\,(\Delta G_{\mathrm{self}}(S') - \Delta G_l)\bigr)\; \\
&\sigma\!\bigl(-\,k_u\,(\Delta G_{\mathrm{self}}(S') - \Delta G_u)\bigr).
\end{aligned}
\end{equation}

where $\Delta G_{\mathrm{self}}(S')$ is the minimum free energy of the
aptamer's intramolecular self-folding, estimated using nearest-neighbour
thermodynamic parameters~\citep{santalucia1998unified} at
$37\,{}^{\circ}\mathrm{C}$, and $\sigma(x) = 1/(1+e^{-x})$ is the logistic
sigmoid.
The product of the two sigmoids forms a soft window over
$\Delta G_{\mathrm{self}}$: when $\Delta G_l < \Delta G_{\mathrm{self}} <
\Delta G_u$, both sigmoids are close to~$1$ and
$\Ephys \approx 0$; outside this range one sigmoid drops toward
zero, pushing $\Ephys$ toward~$1$.
The parameters $\Delta G_l$ and $\Delta G_u$ define the boundaries of
the favourable region, and $k_l$, $k_u$ control how sharply the penalty
rises on each side.

The boundary parameters
($\Delta G_l = -10\;\mathrm{kcal\,mol^{-1}}$,
$\Delta G_u = -2\;\mathrm{kcal\,mol^{-1}}$)
and steepness ($k_l = k_u = 0.75$) are derived from the empirical
distribution of nearest-neighbour self-folding free energies across
random DNA sequences (Appendix~\ref{app:param_derivation}).

\paragraph{Integration into PH-UCB selection.}
This penalty enters the PH-UCB selection rule as a multiplicative
factor $\exp\!\bigl(-\wphys\,\Ephys\bigr)$,
which reduces the UCB score of candidates at either extreme.
The weight $\wphys = 0.1$ makes the penalty a soft preference rather than
a hard filter. Like the consensus prior and uncertainty bonuses, $\Ephys$ is
computed once when a node is expanded and cached for reuse during
selection, adding no extra cost to the selection phase.
$\Delta G_{\mathrm{self}}$ is computed from the sequence alone in
$<\!1\,\mathrm{ms}$, without requiring an additional structure prediction call.

For aptamer types where nearest-neighbour thermodynamic models do not
apply (e.g., G-quadruplex sequences governed by Hoogsteen base-pairing),
we set $\Ephys = 0.5$ (a neutral value that neither penalises nor
rewards the candidate).

Overall, the exploration term is large for actions that are
(i)~plausible under expert consensus,
(ii)~informative due to residual sequence/structure uncertainty, and
(iii)~physically reasonable under task-specific biophysical constraints,
while being down-weighted for candidates at either biophysical extreme.

\subsection{Parameter derivation.}
\label{app:param_derivation}

We use the same calibration principle for all parameters: each sigmoid
should reach its sensitive region (input $\approx \pm 3$) at the most
extreme $\Delta G_{\mathrm{self}}$ values observed in practice.
To ground this in data, we computed $\Delta G_{\mathrm{self}}$ for a
sample of random DNA sequences across the 20--80~nt range using
nearest-neighbour thermodynamics (\texttt{seqfold},
\citealt{santalucia1998unified}, $37\,{}^{\circ}\mathrm{C}$).
The observed distribution spans approximately $-14$ to
$+2\;\mathrm{kcal\,mol^{-1}}$, with a median near
$-2.5\;\mathrm{kcal\,mol^{-1}}$.

For the upper boundary, we set $\Delta G_u = -2\;\mathrm{kcal\,mol^{-1}}$.
A minimal stable hairpin (4-bp stem with a tetraloop) has
$\Delta G \approx -2$ to $-3\;\mathrm{kcal\,mol^{-1}}$ under
nearest-neighbour parameters~\citep{santalucia1998unified};
sequences above this threshold cannot maintain even one stable
secondary-structure element.

For the lower boundary, we set $\Delta G_l = -10\;\mathrm{kcal\,mol^{-1}}$,
corresponding to the 1st--5th percentile of the observed distribution.
Sequences below this value carry multiple stable stem-loops whose
cumulative unfolding cost substantially reduces effective binding
affinity.

The steepness parameters follow from the same calibration.
On the over-stability side, the observed extreme
$\Delta G_{\min} \approx -14\;\mathrm{kcal\,mol^{-1}}$ lies
$4\;\mathrm{kcal\,mol^{-1}}$ below $\Delta G_l$,
giving $k_l = 3\,/\,4 = 0.75$.
On the under-structure side, the observed extreme
$\Delta G_{\max} \approx +2\;\mathrm{kcal\,mol^{-1}}$ lies
$4\;\mathrm{kcal\,mol^{-1}}$ above $\Delta G_u$,
giving $k_u = 3\,/\,4 = 0.75$.
The two sides yield the same steepness, so we use a single
$k_l = k_u = 0.75$; the transition width of $4.4\,/\,0.75
\approx 6\;\mathrm{kcal\,mol^{-1}}$ is intentionally gradual,
reflecting uncertainty in the exact threshold values.

\subsection{Additional details: multi-modality representation and expert dispatch}
\label{sec:app:multimodal}

\algname operates across diverse biomolecular modalities---proteins, peptides, DNA/RNA (single- and double-stranded),
small molecules, ions, and post-translational modifications---without modality-specific fine-tuning or architectural
changes. This generality follows from two complementary design choices: \emph{(i)} a unified all-atom representation
in the folding expert, and \emph{(ii)} a modular, modality-aware dispatch in the inverse-folding step. Across all
settings, the planner state remains $s=(S,X)$ under conditioning $C$ (partners, ligands, motifs, etc.), and the MCTS
logic (selection/expansion/caching/backup) is unchanged.

\paragraph{Unified all-atom tokenization.}
We use Boltz-2 as the folding/refolding expert, whose unified all-atom representation supports proteins, nucleic acids,
small molecules, ions, and covalent modifications within a single framework~\citep{passaro2025boltz2}. As a result,
\algname inherits broad modality coverage “for free”: modality differences are expressed through conditioning $C$ and
task-specific scoring/feasibility cues, rather than through retraining or specialized architectures.

\paragraph{Modality-aware expert dispatch.}
Sequence redesign is implemented as a lightweight dispatch based on the design target type. Concretely, the planner
invokes an appropriate inverse-folding expert for the currently designable component (e.g., ProteinMPNN for proteins
and peptides~\citep{dauparas2022proteinmpnn}, NA-MPNN for nucleic acids~\citep{kubaney2025rna}, and LigandMPNN for
ligand-pocket-aware protein redesign). The state representation, tree policy, and value computation remain fixed;
only the inverse-folding call differs. Table~\ref{tab:app_dispatch} summarizes the dispatch and typical plug-in
biophysical cues used for controllability.

\begin{table}[t]
\centering
\small
\setlength{\tabcolsep}{4pt}
\renewcommand{\arraystretch}{1.05}
\begin{adjustbox}{max width=\columnwidth}
\begin{tabular}{l l l l}
\toprule
\textbf{Setting} & \textbf{Designable component} & \textbf{Inverse-folding expert} & \textbf{Typical biophysics cue} \\
\midrule
Protein/peptide design & protein/peptide sequence & ProteinMPNN & clash/packing/contact \\
Ligand-conditioned protein design & pocketed protein sequence & LigandMPNN & sterics / pose geometry \\
RNA/DNA aptamer design & NA sequence & NA-MPNN & $\Delta G_{\mathrm{self}}$, base-pair potentials \\
Hybrid complexes & per-chain (as applicable) & dispatched & task-specific \\
\bottomrule
\end{tabular}
\end{adjustbox}
\caption{\textbf{Modality-aware inverse-folding dispatch.} The folding expert (Boltz-2) is shared across settings via a unified all-atom representation; modality differences enter primarily through the inverse-folding backend and optional plug-in biophysical feasibility cues.}
\label{tab:app_dispatch}
\end{table}

\subsection{Inference-time structural guidance for nucleic acids}
\label{sec:app:struct_guidance}
Sequence-level biophysical signals---such as self-folding free energy
or base-pair assignments---capture important physical constraints, but
they do not directly influence the structure generation process.
A more direct form of biophysical control is to inject physical
constraints \emph{inside} the structure generation itself, steering the
diffusion model's denoising trajectory toward conformations that are
consistent with the sequence's known secondary-structure preferences.

For a candidate sequence $S'$ with self-folding stability in the
thermodynamically favourable range
($\Delta G_l \le \Delta G_{\mathrm{self}}(S') \le \Delta G_u$;
boundaries derived in Appendix~\ref{app:param_derivation}),
we derive an inference-time structural guidance signal from the
\texttt{seqfold} computation.

\paragraph{\texttt{seqfold} outputs.}
The \texttt{seqfold} computation for a candidate sequence $S'$ produces
two outputs from a single call ($<\!1\,\mathrm{ms}$):
\begin{enumerate}
  \item $\Delta G_{\mathrm{self}}(S')$ --- the minimum free energy of
        intramolecular self-folding.
  \item $P(S')$ --- the MFE base-pair assignment, i.e., the set of
        intramolecular base pairs predicted to form in the free aptamer.
\end{enumerate}
This section uses $P(S')$ as a \emph{contact guidance potential} inside
Boltz-2's structure prediction (described below).
The contact guidance adds no extra computational overhead because
$P(S')$ is already available from the same \texttt{seqfold} call that
produces $\Delta G_{\mathrm{self}}$.

\paragraph{Inference-time structural guidance via Boltz-2 potentials.}
We activate Boltz-2's inference-time potential-energy steering
(\texttt{use\_potentials}~$=$~\texttt{True}), which steers the denoising
trajectory via gradient-based $\hat{x}_0$ correction and Feynman-Kac
particle resampling~\citep{del2006sequential} entirely within the forward
pass---no backpropagation through the diffusion model is required.

Within this steering framework, we inject $P(S')$ as a \emph{contact
guidance potential}: aptamer positions $I$ that form contacts with the
protein interface in the current hallucinated complex $X'$ are identified,
and base pairs in $P(S')$ whose both endpoints lie \emph{outside} $I$
(non-interface pairs) are used as soft structural constraints.
This encourages the aptamer scaffold to retain its pre-organized secondary
structure away from the binding site, while bases at the interface
($\in I$) are left unconstrained so they remain available for protein
contact.
The contact weight is higher during the high-noise structural assembly
phase and disabled near convergence, following Boltz-2's default schedule.

Formally, let $\mathcal{P}_{\text{free}} = \{(i,j) \in P(S') : i \notin I,\, j \notin I\}$
denote the set of non-interface base pairs predicted by \texttt{seqfold}.
The guidance potential is
\begin{equation}
U_{\text{contact}}(X) \;=\; \sum_{(i,j)\,\in\,\mathcal{P}_{\text{free}}}
w_{ij}\,\bigl(d_{ij}(X) - d_0\bigr)^2,
\label{eq:contact_potential}
\end{equation}
where $d_{ij}(X)$ is the distance between the heavy atoms of positions $i$
and $j$ in structure $X$, $d_0$ is the target contact distance for a
Watson-Crick base pair, and $w_{ij}$ is a weight that decays with
diffusion noise level.
This potential is incorporated into the Boltz-2 denoising step via the
$\hat{x}_0$-correction mechanism, yielding a guided reverse transition.

\subsection{Reproducibility and Hyperparameters}
\label{app:hyperparams}

Table~\ref{tab:hyperparams_main} summarizes the audited configuration used for the main experiments. Unless otherwise stated, the same configuration is used across RNA aptamer, DNA aptamer, protein--protein binder, and ligand-conditioned design tasks. We do not tune search hyperparameters per dataset; modality differences enter only through the inverse-folding backend and the task conditioning passed to the folding expert.

\begin{table}[h]
\centering
\caption{Main hyperparameters and configuration.}
\label{tab:hyperparams_main}
\small
\begin{tabularx}{\linewidth}{>{\raggedright\arraybackslash}p{0.22\linewidth}>{\raggedright\arraybackslash}p{0.30\linewidth}X}
\toprule
Component & Parameter & Value \\
\midrule
Tree search & Fold-call budget $T$ & 100 \\
& Expansion width $K$ & 4 \\
& Maximum tree depth $D_{\max}$ & 5 \\
& Maximum MCTS iterations & 50 \\
& Fold calls per iteration & 8 \\
& Early-stop patience & 10 iterations without improvement \\
& Exploration constant $c_p$ & $\sqrt{2}\approx 1.414$ \\
& Backup rule & sum-backup, $Q=\texttt{total\_reward}/\texttt{visits}$ \\
& Main-table seed & 42 \\
& Ablation seeds & $\{11,23,47\}$ \\
\midrule
Value function & $w_{\mathrm{conf}}$ & 1.0 \\
& $w_{\mathrm{iface}}$ & 0.5 \\
& $w_{\mathrm{motif}}$ & 5.0 when motif constraints are specified; 0 otherwise \\
\midrule
Folding expert & Model & Boltz-2 \\
& Recycling steps & 3 \\
& Diffusion denoising steps & 200 \\
& Sampling steps & 200 \\
& Diffusion samples per call & 1 \\
\midrule
Inverse-folding experts & Protein / peptide & ProteinMPNN \\
& Nucleic acid (RNA / DNA) & NA-MPNN \\
& Ligand-pocket-aware protein & LigandMPNN \\
& Sampling temperature $\tau$ & 0.1 \\
& Samples per inverse-fold call & 4 \\
\midrule
Compute & GPU & NVIDIA A100 80GB \\
& Per target / seed & 1 GPU \\
\bottomrule
\end{tabularx}
\end{table}

For each target, we report the design with the highest ipTM among all states evaluated during search, including both internal and leaf states. This candidate-selection rule is fixed across all reported MCTH runs.

%% file: supp_additional_experiments.tex
\section{Additional Experiments}
\label{sec:app:additional_experiments}

This supplement reports additional evaluations showing that the same \algname pipeline can be instantiated across
modalities and constraint types beyond the three main-text benchmarks (Sec.~\ref{sec:experiments}).
\fix{For concision, we currently include only protein--ligand binding and the corresponding inverse-folding ablation
(Appendix~\ref{sec:app:ligand_ablation}); other additional experiments are commented out below.}

\subsection{Protein--Ligand Binding Design}
\label{sec:app:ligand}

We evaluate \algname on protein--small-molecule binding design, where the objective is to generate a protein sequence that forms a confident complex with a specified ligand.

\paragraph{Benchmark.}
We construct a 10-target ligand-binding benchmark from four PDB complexes: 5SDV (IAI; chains A/B/C/D), 7BKC (FAD; chains A/a/E/e),
7C7M (SAM; chain A), and 7V11 (OQO; chain A). Each target specifies a fixed ligand (CCD code) and a binding protein chain.

\paragraph{Setup.}
\algname instantiates MCTS with Boltz-2 folding in the loop and \textbf{LigandMPNN} as the inverse-folding proposal model. Starting from random sequences,
the search iterates folding, scoring with an ipTM-dominant value function, and MCTS expansion with ligand-aware inverse-fold proposals. We use a fixed budget
of up to 100 Boltz-2 fold calls per target.

\paragraph{Baselines and evaluation.}
We compare against (i) RFDiffusion-AA + LigandMPNN and (ii) ODesign (rigid ligand checkpoint). All designs are evaluated under a common
Boltz-2 re-fold protocol. We report \textbf{ipTM} (higher is better). \fix{A specialist comparison to BoltzDesign1 is provided as a set-aside table below.}

\paragraph{Results.}
\algname attains a higher mean ipTM ($0.901$) than RFDiffusion-AA ($0.850$) and ODesign ($0.460$), with the main gap concentrated on the most
challenging FAD cofactor targets (higher pocket-precision demands). Ablations are in Appendix~\ref{sec:app:ligand_ablation}.

\begin{table}[t]
\centering
\small
\setlength{\tabcolsep}{4pt}
\renewcommand{\arraystretch}{1.05}
\begin{adjustbox}{max width=\columnwidth}
\begin{tabular}{l cc c}
\toprule
Target & RFDiff-AA & \algname & ODesign \\
\midrule
5SDV\_IAI\_A & 0.932 & \textbf{0.949} & 0.691 \\
5SDV\_IAI\_B & \textbf{0.959} & 0.939 & 0.516 \\
5SDV\_IAI\_C & \textbf{0.924} & 0.895 & 0.269 \\
5SDV\_IAI\_D & \textbf{0.961} & 0.902 & 0.449 \\
7BKC\_FAD\_A & 0.465 & \textbf{0.848} & 0.200 \\
7BKC\_FAD\_a & 0.878 & \textbf{0.919} & 0.291 \\
7BKC\_FAD\_E & \textbf{0.839} & 0.833 & 0.410 \\
7BKC\_FAD\_e & 0.720 & \textbf{0.846} & 0.510 \\
7C7M\_SAM\_A & 0.928 & \textbf{0.940} & 0.847 \\
7V11\_OQO\_A & 0.897 & \textbf{0.938} & 0.413 \\
\midrule
Mean & 0.850 & \textbf{0.901} & 0.460 \\
\bottomrule
\end{tabular}
\end{adjustbox}
\caption{\textbf{Protein--ligand binding under Boltz-2 refold evaluation.} Best ipTM per target (higher is better). Bold indicates the better of \algname and RFDiff-AA for each target; \algname attains a higher mean ipTM.}
\label{tab:ligand_iptm_summary_main}
\end{table}

\section{Additional Ablations}
\label{sec:app:ligand_ablation}

\subsection{Ligand-aware vs.\ ligand-blind inverse folding}
\label{sec:app:ligand_ligandmpnn_ablation}

To isolate the effect of ligand-aware sequence redesign, we compare \algname under two inverse-folding backends:
(i) \textbf{LigandMPNN} (ligand-aware; conditions on ligand atoms in the binding pocket), and
(ii) \textbf{ProteinMPNN} (ligand-blind; conditions only on protein backbone atoms).
All other components are held fixed: identical MCTS configuration, identical ipTM-dominant scoring for ligand tasks,
and the same Boltz-2 evaluation protocol as in Appendix~\ref{sec:app:ligand}.
We report best-of-$N$ ipTM (higher is better) and iPDE (lower is better).

\paragraph{Summary.}
Ligand-aware inverse folding substantially improves both interface confidence and interface precision:
LigandMPNN improves mean ipTM from 0.826 to 0.901 and reduces mean iPDE from 4.20 to 3.02.

\begin{table}[t]
\centering
\small
\setlength{\tabcolsep}{4pt}
\renewcommand{\arraystretch}{1.05}
\begin{adjustbox}{max width=\columnwidth}
\begin{tabular}{lccccc}
\toprule
& \multicolumn{2}{c}{\textbf{ipTM} $\uparrow$} && \multicolumn{2}{c}{\textbf{iPDE} $\downarrow$} \\
\cmidrule(lr){2-3}\cmidrule(lr){5-6}
\textbf{Ligand} & LigandMPNN & ProteinMPNN && LigandMPNN & ProteinMPNN \\
\midrule
IAI (4 targets) & \textbf{0.921} & 0.900 && 1.60 & \textbf{1.18} \\
FAD (4 targets) & \textbf{0.862} & 0.730 && \textbf{5.46} & 7.98 \\
SAM (1 target)  & \textbf{0.940} & 0.818 && \textbf{0.92} & 4.55 \\
OQO (1 target)  & \textbf{0.938} & 0.924 && 1.02 & \textbf{0.84} \\
\bottomrule
\end{tabular}
\end{adjustbox}
\caption{\textbf{Ablation by ligand type.} Ligand-aware inverse folding yields the largest improvements on FAD and SAM, where pocket complementarity is most critical.}
\label{tab:ligand_ablation_by_ligand}
\end{table}

\paragraph{Per-target results.}
\begin{table}[t]
\centering
\small
\setlength{\tabcolsep}{4pt}
\renewcommand{\arraystretch}{1.05}
\begin{adjustbox}{max width=\columnwidth}
\begin{tabular}{l cc cc}
\toprule
& \multicolumn{2}{c}{\textbf{ipTM} $\uparrow$} & \multicolumn{2}{c}{\textbf{iPDE} $\downarrow$} \\
\cmidrule(lr){2-3}\cmidrule(lr){4-5}
\textbf{Target} & LigandMPNN & ProteinMPNN & LigandMPNN & ProteinMPNN \\
\midrule
5SDV\_IAI\_A & 0.949 & 0.889 & 2.24 & 1.25 \\
5SDV\_IAI\_B & 0.939 & 0.921 & 1.28 & 1.00 \\
5SDV\_IAI\_C & 0.895 & 0.876 & 1.65 & 1.42 \\
5SDV\_IAI\_D & 0.902 & 0.914 & 1.23 & 1.04 \\
7BKC\_FAD\_A & 0.848 & 0.733 & 7.69 & 9.49 \\
7BKC\_FAD\_a & 0.919 & 0.737 & 5.38 & 8.46 \\
7BKC\_FAD\_E & 0.833 & 0.753 & 6.60 & 7.75 \\
7BKC\_FAD\_e & 0.846 & 0.695 & 2.17 & 6.22 \\
7C7M\_SAM\_A & 0.940 & 0.818 & 0.92 & 4.55 \\
7V11\_OQO\_A & 0.938 & 0.924 & 1.02 & 0.84 \\
\bottomrule
\end{tabular}
\end{adjustbox}
\caption{\textbf{Per-target ablation results.} Ligand-aware inverse folding improves both ipTM and iPDE on most targets, with the strongest gains on FAD and SAM.}
\label{tab:ligand_ablation_per_target}
\end{table}

\subsection{Evaluation Protocol and Cross-Oracle Validation}
\label{app:evaluation_protocol}

We distinguish between \emph{search-time} and \emph{held-out} evaluation.

\paragraph{Search-time oracle evaluation.}
For each MCTH instantiation, candidate states are scored using the folding-model
signals available to the search procedure. The modality-spanning main experiments
use Boltz-2 as the search-time folding oracle. For comparisons that can be
re-scored consistently, including ODesign, RFDiffusion, RFDiffusion-AA, and
Complexa, generated sequences are evaluated using the same Boltz-2 refolding
protocol. BoltzDesign1 instead uses its native Boltz evaluation during optimization;
we therefore report it separately while matching the number of folding/evaluator
calls where applicable.

These scores measure optimization under a common computational proxy and should
not be interpreted as oracle-independent evidence of physical binding.

\paragraph{Matched-budget evaluation.}
To separate search quality from raw access to the evaluator, we match candidate or
folding-call budgets whenever possible. In the main benchmarks, ODesign,
RNAFrameFlow, Complexa, and BoltzDesign1 are compared at up to 100 candidate or
folding/evaluator calls where applicable. Appendix~\ref{app:matched_budget}
further compares MCTH against exploration-off search, greedy cycling, and
best-of-$N$ sampling under identical initialization, proposal models, folding
oracle, and evaluation budget.

\paragraph{Held-out predictor evaluation.}
We additionally evaluate designs with structure predictors that are not used to
guide the corresponding search. AlphaFold3 is used to re-score the RNA and DNA
aptamer benchmarks. For the protein-binder dual-expert experiment, Boltz-2 and
AlphaFold2 are used during search, while Chai-1 is held out entirely from proposal,
tree selection, and output ranking. These evaluations test whether improvements
transfer beyond the optimized predictor rather than merely reproducing its scoring
preferences.

\paragraph{Oracle-independent physical evaluation.}
For protein binders, we additionally use Rosetta InterfaceAnalyzer as a
non-search evaluation of interface geometry and energetics. We report
size-normalized interface energy together with packing and buried unsatisfied
hydrogen-bond statistics. These metrics provide an additional physical-plausibility
check but are not treated as experimental affinity measurements.

\paragraph{Final-design pass rate.}
Best-per-target ipTM measures whether a method can discover a strong candidate but
does not characterize design yield. We therefore also compare final protein-binder
archives under a common held-out binder-quality filter. Candidates are selected
without access to held-out metrics, and pass rates are computed only after final
archive construction.

\subsection{Matched-Budget Search Ablations}
\label{app:matched_budget_ablation}

To isolate the effect of search-guided compute allocation, we run matched-budget ablations under the same 100 folding-call budget. For each target, all variants use identical initialization, evaluation, scoring, folding oracle, and inverse-folding backend. The only difference is the planning strategy.

We compare the following variants:
\begin{itemize}
    \item \textbf{MCTH (full):} tree search with exploration.
    \item \textbf{Exploration-off} $(c_p=0)$: the same tree search without the exploration term, corresponding to pure exploitation.
    \item \textbf{Greedy cycling:} sequential inverse-folding $\rightarrow$ folding refinement along a single trajectory, without branching.
    \item \textbf{Best-of-$N$:} independent sampling under the same fold-call budget, without feedback or iterative refinement.
\end{itemize}

Greedy cycling isolates the effect of iterative refinement without branching, while best-of-$N$ isolates pure sampling from strong pretrained proposal models without feedback. Together, these baselines test whether MCTH's gains come from search-guided compute allocation rather than refinement or sampling alone.

\begin{table}[h]
\centering
\caption{Matched-budget ablation on RNA aptamer design. All values are mean $\pm$ standard deviation over three seeds.}
\label{tab:ablation_rna}
\small
\begin{tabular}{lccccc}
\toprule
Variant & 7WKP A$\rightarrow$B & 7YEW A$\rightarrow$B & 8GXB C$\rightarrow$B & Mean & $\Delta$ vs. Full \\
\midrule
\textbf{MCTH (full)} & \textbf{0.902$\pm$0.010} & \textbf{0.979$\pm$0.002} & \textbf{0.858$\pm$0.006} & \textbf{0.913} & 0.000 \\
Exploration-off $(c_p=0)$ & 0.756$\pm$0.007 & 0.972$\pm$0.002 & 0.791$\pm$0.040 & 0.839 & -0.073 \\
Greedy cycling & 0.750$\pm$0.004 & 0.957$\pm$0.003 & 0.646$\pm$0.012 & 0.784 & -0.129 \\
Best-of-$N$ random & 0.752$\pm$0.030 & 0.953$\pm$0.001 & 0.629$\pm$0.010 & 0.778 & -0.135 \\
\bottomrule
\end{tabular}
\end{table}

\begin{table}[h]
\centering
\caption{Matched-budget ablation on DNA aptamer design. All values are mean $\pm$ standard deviation over three seeds.}
\label{tab:ablation_dna}
\small
\begin{tabular}{lccccc}
\toprule
Variant & 7XVN A\_N & 7YUK A\_C & 8PMF A\_B & Mean & $\Delta$ vs. Full \\
\midrule
\textbf{MCTH (full)} & \textbf{0.955$\pm$0.002} & \textbf{0.946$\pm$0.008} & \textbf{0.964$\pm$0.005} & \textbf{0.955} & 0.000 \\
Exploration-off $(c_p=0)$ & 0.909$\pm$0.011 & 0.904$\pm$0.009 & 0.930$\pm$0.009 & 0.914 & -0.041 \\
Greedy cycling & 0.756$\pm$0.014 & 0.909$\pm$0.018 & 0.819$\pm$0.006 & 0.828 & -0.127 \\
Best-of-$N$ random & 0.753$\pm$0.010 & 0.900$\pm$0.006 & 0.834$\pm$0.005 & 0.829 & -0.126 \\
\bottomrule
\end{tabular}
\end{table}

\begin{table}[h]
\centering
\caption{Matched-budget ablation on protein--protein binder design. All values are mean $\pm$ standard deviation over three seeds.}
\label{tab:ablation_ppi}
\small
\begin{tabular}{lcccc}
\toprule
Variant & 3DI3 & 4O3V & Mean & $\Delta$ vs. Full \\
\midrule
\textbf{MCTH (full)} & \textbf{0.912$\pm$0.002} & \textbf{0.964$\pm$0.002} & \textbf{0.938} & 0.000 \\
Exploration-off $(c_p=0)$ & 0.897$\pm$0.008 & 0.954$\pm$0.016 & 0.925 & -0.013 \\
Greedy cycling & 0.901$\pm$0.023 & 0.955$\pm$0.013 & 0.928 & -0.011 \\
Best-of-$N$ random & 0.803$\pm$0.005 & 0.900$\pm$0.008 & 0.851 & -0.087 \\
\bottomrule
\end{tabular}
\end{table}

\begin{table}[h]
\centering
\caption{Summary of matched-budget ablations across all targets.}
\label{tab:ablation_summary}
\small
\begin{tabular}{lcccc}
\toprule
Variant & Overall mean & Protein mean & RNA mean & DNA mean \\
\midrule
\textbf{MCTH (full)} & \textbf{0.935} & \textbf{0.938} & \textbf{0.913} & \textbf{0.955} \\
Exploration-off $(c_p=0)$ & 0.889 & 0.925 & 0.839 & 0.914 \\
Greedy cycling & 0.836 & 0.928 & 0.784 & 0.828 \\
Best-of-$N$ random & 0.815 & 0.851 & 0.778 & 0.829 \\
\bottomrule
\end{tabular}
\end{table}

Across all targets, MCTH achieves the highest final quality, with especially large gains on RNA and DNA tasks. This suggests that the improvement comes from search-guided compute allocation rather than simply from iterative refinement or a larger independent candidate pool.

\subsection{Effective Convergence Diagnostic}
\label{app:convergence}

Final performance alone does not reveal whether additional folding calls continue to produce
useful search progress. We therefore measure the last folding call at which the incumbent best
score improves by more than $\epsilon=0.002$. Runs with no improvement above this threshold
are assigned a last-improvement call of zero and counted as immediately stagnant.

\begin{table}[t]
\centering
\small
\setlength{\tabcolsep}{6pt}
\begin{tabular}{lccc}
\toprule
Variant & Final mean & Never improve & Avg. last improvement \\
\midrule
MCTH (full) & \textbf{0.935} & \textbf{0.0\%} & \textbf{46.6} \\
Exploration-off ($c_p=0$) & 0.889 & 61.9\% & 29.3 \\
Greedy cycling & 0.836 & 53.8\% & 23.5 \\
Best-of-$N$ & 0.815 & 44.4\% & 33.3 \\
\bottomrule
\end{tabular}
\caption{
Effective convergence at $\epsilon=0.002$ under the matched folding-call budget.
``Never improve'' denotes runs with no improvement above $\epsilon$ after initialization.
}
\label{tab:effective_convergence}
\end{table}

MCTH continues to extract useful improvements later in the evaluation budget: every MCTH run
improves beyond the threshold, with the last improvement occurring at fold call 46.6 on average.
In contrast, 44--62\% of the non-adaptive or reduced-search runs never make a nontrivial
improvement after initialization. Together with the matched-budget results in
Appendix~\ref{app:matched_budget}, this indicates that the advantage of MCTH is not simply
additional iterative refinement; tree search changes how useful evaluations are allocated over the
course of optimization.

\subsection{Cross-Oracle Agreement Analysis}
\label{app:cross_oracle_agreement}

Held-out re-scoring tests whether improvements under the search-time oracle transfer to an
independent structure predictor. We further quantify this relationship at the \emph{design level}:
given a fixed target, do Boltz-2 and the held-out predictor rank candidate designs similarly?

\paragraph{Design-level agreement.}
We compute within-target Spearman correlation between Boltz-2 and AlphaFold3 scores and
bootstrap over targets to obtain 95\% confidence intervals. This within-target analysis removes
between-target difficulty effects that can otherwise inflate pooled correlations.

\begin{table}[t]
\centering
\small
\setlength{\tabcolsep}{6pt}
\begin{tabular}{lccc}
\toprule
Modality & $n$ designs & Within-target $\rho$ & 95\% CI \\
\midrule
DNA aptamer & 50  & 0.487 & $[0.29,\;0.67]$ \\
Protein binder & 255 & 0.252 & $[0.05,\;0.46]$ \\
RNA aptamer & 77  & 0.173 & $[-0.06,\;0.36]$ \\
\bottomrule
\end{tabular}
\caption{
Design-level agreement between Boltz-2 and AlphaFold3 interface scores.
Correlations are computed within target and confidence intervals are obtained by
bootstrap over targets.
}
\label{tab:cross_oracle_iptm}
\end{table}

Agreement is moderate for DNA aptamers, weak but positive for protein binders, and not
distinguishable from zero for RNA aptamers. Importantly, the pooled raw ipTM correlation
across all designs is much larger ($\rho=0.670$) than the within-target correlation
($\rho=0.261$). The pooled value is therefore driven substantially by target-level difficulty
that both predictors capture, rather than by agreement over which design is best for a fixed target.

\paragraph{Global folding versus interface ranking.}
The disagreement is specific to interface-level discrimination rather than global folding quality.
For protein binders ($n=255$), within-target agreement in pLDDT is
$\rho=0.762$ (95\% CI $[0.64,\;0.85]$), compared with only
$\rho=0.252$ (95\% CI $[0.05,\;0.46]$) for ipTM.
The non-overlapping confidence intervals indicate that the two predictors agree substantially more
on whether a complex is globally well folded than on the relative quality of its binding interface.
This distinction is important for binder and aptamer design, where interface discrimination is the
primary optimization objective.

\paragraph{Use as a screening proxy.}
We next ask whether Boltz-2 can serve as a practical screening proxy for selecting designs that
score highly under the held-out oracle. For each target, we measure whether the Boltz-selected
top design falls in the held-out top set (top-1 hit) and the precision of the Boltz top quartile
with respect to the held-out top quartile.

\begin{table}[t]
\centering
\small
\setlength{\tabcolsep}{5pt}
\begin{tabular}{lcccc}
\toprule
Modality
& Top-1 hit
& Random
& Prec.@25\%
& Random \\
\midrule
DNA aptamer     & 0.600 & 0.200 & 0.600 & 0.200 \\
Protein binder  & 0.167 & 0.053 & 0.427 & 0.241 \\
RNA aptamer     & 0.286 & 0.091 & 0.143 & 0.273 \\
\bottomrule
\end{tabular}
\caption{
Screening utility of Boltz-2 for selecting designs favored by the held-out predictor.
Random columns give the corresponding chance-level expectation under the same target-level
candidate-set sizes.
}
\label{tab:cross_oracle_screening}
\end{table}

Boltz-2 is a useful but imperfect proxy for DNA, where both top-1 recovery and top-quartile
precision are approximately three times their random baselines. For protein binders, screening
performance is above random but substantially weaker. For RNA, the observed top-quartile
precision does not exceed random expectation.

\paragraph{RNA score saturation.}
The weak RNA agreement is partly explained by restricted score dynamic range rather than clear
opposite rankings. Across the RNA candidates, the ipTM interquartile range is only $0.080$ under
AlphaFold3 and $0.022$ under Boltz-2; the latter has a median ipTM of approximately $0.946$.
Thus, many RNA designs lie near the Boltz-2 score ceiling, leaving little resolution for
within-target ranking.

\paragraph{Method-level rankings.}
At the method level, the two predictors preserve the broad direction of performance but are not
equivalent. Kendall's $\tau$ between method rankings is $0.333$ for DNA (10 targets),
$0.738$ for protein binders (5 common targets), and $0.333$ for RNA (7 targets).
These results motivate reporting held-out evaluation alongside the search-time oracle rather than
treating either predictor as a ground-truth interface metric.

\paragraph{Exploratory check for search-oracle preference.}
Finally, we test whether methods that optimize Boltz-derived feedback tend to receive relatively
better rankings under Boltz-2 than under AlphaFold3. Across target--method pairs, Boltz-based
systems shift by $+0.367$ rank positions on average (positive means relatively favored by Boltz-2),
whereas non-Boltz systems shift by $-0.353$, a difference of $0.720$ rank positions.
A 20,000-permutation test that shuffles the Boltz/non-Boltz label within modality gives
$p=0.067$. We therefore treat this as a suggestive diagnostic rather than evidence of a
systematic evaluator bias. It nevertheless reinforces the need for the held-out evaluations used
throughout Sec.~4.

\subsection{Dual-Expert Search and Final-Archive Yield}
\label{app:dual_expert_details}

We provide additional implementation and archive-level evaluation details for the dual-expert
protein-binder experiment in Sec.~\ref{sec:exp_ppi}. The purpose of this experiment is to
test whether the multi-expert mechanism defined in Sec.~\ref{sec:uncertainty_mcts} improves
robustness beyond optimization against a single folding predictor.

\paragraph{Dual-expert configuration.}
We retain the same MCTS procedure and ProteinMPNN sequence-proposal mechanism used in the
single-expert protein-binder setting. The difference is that each candidate sequence is evaluated
with both Boltz-2 and AlphaFold2. Their structural-quality signals contribute to node evaluation,
while cross-expert agreement, disagreement, and uncertainty enter the CU-PUCT search rule.
No component model is fine-tuned or differentiated through.

Each Boltz-2 or AlphaFold2 structure-prediction invocation is counted as one folding/evaluator
call. Thus, evaluating one sequence with both experts consumes two calls from the fixed inference
budget. This accounting prevents the dual-expert variant from receiving uncounted additional
structure-prediction compute.

Chai-1 is not used for sequence proposal, node evaluation, tree selection, backup, or archive
ranking. It is run only after search and therefore serves as a held-out predictor.

\paragraph{Cross-predictor transfer.}
Across the 11 CaoData targets shared by the single-expert, dual-expert, and RFDiffusion
evaluations, the original Boltz-only MCTH obtains mean best ipTM values of
$0.69/0.19/0.18$ under Boltz-2/AF2/Chai-1, respectively. The dual-expert variant obtains
$0.74/0.61/0.57$. Thus, introducing AF2 as a second search-time expert substantially improves
performance under AF2 while maintaining Boltz-2 performance, and the improvement transfers to
the held-out Chai-1 evaluator. Full per-target results are reported in
Table~\ref{tab:dual_expert_cross_oracle}.

For MCTH, the evaluator-wise values in Table~\ref{tab:dual_expert_cross_oracle} are maxima
over three retained final designs. Because the maximum is taken independently for each evaluator,
the Boltz-2, AF2, and Chai-1 values for a target need not correspond to the same sequence.
RFDiffusion values are maxima over ten outputs, so we do not interpret the unequal-pool
best-of-output comparison as a direct design-yield comparison.

\paragraph{Final-archive construction.}
To measure yield beyond a single best candidate, we reconstruct a reward-ranked archive from the
dual-expert search logs. Within each target, candidates from the available dual-expert runs are
pooled and deduplicated by exact binder amino-acid sequence identity after removing whitespace
and converting sequences to uppercase. Candidate ranking uses only the joint reward recorded
during search; no held-out Chai-1 quantity is used for retrospective selection.

We report the top-$K$ unique archive for $K\in\{5,10,25\}$. Intermediate MCTS nodes are
optimization states rather than independently submitted final designs and are therefore not used
as the denominator of the final-design pass rate.

\paragraph{Held-out binder-quality filter.}
After archive construction, each candidate is independently evaluated with Chai-1.
We define a design as passing when its held-out interface confidence satisfies
$\mathrm{ipTM}_{\mathrm{Chai}}\geq 0.50$.
This threshold is fixed before method comparison and is applied identically to
\algname and RFDiffusion. Chai-1 is never used during sequence proposal, MCTS
selection, reward computation, or archive ranking, so the pass criterion is fully
out of the search loop. All designs in the reported archives have valid Chai-1
evaluations; missing or failed evaluations, if present, are excluded from the valid
denominator and reported separately.

\begin{table}[t]
\centering
\small
\setlength{\tabcolsep}{5pt}
\begin{tabular}{lccc}
\toprule
Archive & Passing/valid & Pass rate & Targets $\geq 1$ pass \\
\midrule
Dual-expert MCTH, top-5  & 27/55  & \textbf{49.1\%} & 8/11 \\
Dual-expert MCTH, top-10 & 41/110 & \textbf{37.3\%} & 8/11 \\
Dual-expert MCTH, top-25 & 84/275 & \textbf{30.5\%} & 9/11 \\
\bottomrule
\end{tabular}
\caption{
Final-archive yield for dual-expert MCTH.
Candidates are ranked only by the in-loop joint reward before applying the held-out filter.
}
\label{tab:dual_expert_archive}
\end{table}

The pass fraction decreases as progressively lower-ranked candidates enter the archive
($49.1\%\rightarrow37.3\%\rightarrow30.5\%$), while target coverage increases from
8/11 to 9/11. This behavior is consistent with a precision--coverage trade-off rather than a
single threshold-sensitive result.

\paragraph{Matched comparison with RFDiffusion.}
For a direct baseline comparison, we use exactly ten final designs per target for both methods on
the same 11-target set and apply the identical held-out filter. MCTH uses its reward-ranked
top-10 unique archive, while RFDiffusion uses its ten generated outputs without held-out
re-ranking.

\begin{table}[t]
\centering
\small
\setlength{\tabcolsep}{5pt}
\begin{tabular}{lcccc}
\toprule
Method & Outputs/target & Passing/valid & Pass rate & Targets $\geq1$ pass \\
\midrule
Dual-expert MCTH & 10 & 41/110 & \textbf{37.3\%} & 8/11 \\
RFDiffusion      & 10 & 22/110 & 20.0\%          & 8/11 \\
\bottomrule
\end{tabular}
\caption{
Matched final-design yield using ten outputs per target and the same held-out
binder-quality filter.
}
\label{tab:dual_expert_passrate_appendix}
\end{table}

Under equal output counts, dual-expert MCTH produces 41 passing designs compared with 22 for
RFDiffusion, corresponding to a $17.3$ percentage-point or approximately $1.86\times$ higher
pass rate. Both methods identify at least one passing design on 8/11 targets. The result therefore
reflects a higher density of qualifying outputs rather than broader target coverage alone.

\paragraph{RFDiffusion held-out score distribution.}
For completeness, Table~\ref{tab:rfdiff_heldout_details} reports the per-target held-out scores used
in the matched baseline analysis.

\begin{table}[t]
\centering
\small
\setlength{\tabcolsep}{5pt}
\begin{tabular}{lccc}
\toprule
Target & Chai pass/10 & Chai mean/best ipTM & AF2 mean/best ipTM \\
\midrule
1DJS & 1/10 & 0.327 / 0.594 & 0.166 / 0.278 \\
1MOX & 0/10 & 0.252 / 0.498 & 0.148 / 0.240 \\
1XIW & 6/10 & 0.540 / 0.808 & 0.536 / 0.840 \\
2GY7 & 5/10 & 0.519 / 0.729 & 0.419 / 0.669 \\
2IFG & 2/10 & 0.288 / 0.700 & 0.259 / 0.783 \\
3DI3 & 0/10 & 0.228 / 0.258 & 0.233 / 0.616 \\
3FKD & 0/10 & 0.230 / 0.310 & 0.559 / 0.883 \\
3MJG & 1/10 & 0.248 / 0.525 & 0.107 / 0.159 \\
3ZTJ & 1/10 & 0.280 / 0.660 & 0.258 / 0.557 \\
4O3V & 3/10 & 0.451 / 0.780 & 0.349 / 0.751 \\
4OGA & 3/10 & 0.430 / 0.685 & 0.422 / 0.656 \\
\bottomrule
\end{tabular}
\caption{
Per-target RFDiffusion held-out evaluation for the matched top-10 comparison.
Mean/best values are computed over the ten outputs for each target.
}
\label{tab:rfdiff_heldout_details}
\end{table}

Together, the cross-predictor and archive analyses address two distinct questions.
The dual-expert experiment tests whether search guided by more than one structural model transfers
to a predictor outside the optimization loop, while the matched pass-rate analysis tests whether this
robustness extends beyond a single best candidate to the final output distribution. Neither analysis
constitutes experimental binding validation; rather, they measure cross-predictor robustness and
computational final-design yield under controlled evaluation protocols.

\subsection{Oracle-Independent Rosetta Interface Evaluation}
\label{app:rosetta_interface}

Structure-predictor confidence is not a physical energy and may preserve biases of the models used
during search. We therefore additionally evaluate protein-binder designs with Rosetta
InterfaceAnalyzer, which is not used for proposal generation, MCTS selection, archive ranking, or
any other component of MCTH.

We evaluate the six CaoData targets shared by MCTH and RFDiffusion and the five with available
BindCraft designs. We report interface separation energy $\Delta G_{\mathrm{sep}}$, the same energy
normalized by buried interface area, packing quality (packstat), and buried unsatisfied hydrogen
bonds.

\begin{table}[t]
\centering
\small
\setlength{\tabcolsep}{5pt}
\begin{tabular}{lccccc}
\toprule
Method & $n$ & $\Delta G_{\mathrm{sep}}$ & $\Delta G/100\,\text{\AA}^2$ &
Packstat & $\Delta$unsat H-bonds \\
\midrule
MCTH        & 6 & -71.2 & -3.12 & 0.577 & 5.7 \\
BindCraft   & 5 & -65.7 & \textbf{-3.40} & 0.600 & 5.2 \\
RFDiffusion & 6 & -54.2 & -3.13 & \textbf{0.602} & 6.5 \\
\bottomrule
\end{tabular}
\caption{
Oracle-independent Rosetta InterfaceAnalyzer evaluation of protein-binder designs.
More negative interface energy and higher packstat are favorable; fewer buried unsatisfied
hydrogen bonds are preferred.
}
\label{tab:rosetta_interface}
\end{table}

Raw $\Delta G_{\mathrm{sep}}$ is sensitive to interface size, so we treat the area-normalized energy
as the more interpretable comparison. Under this measure, MCTH is essentially matched with
RFDiffusion ($-3.12$ vs.\ $-3.13$ per $100\,\text{\AA}^2$), while BindCraft is somewhat more
favorable ($-3.40$). Packing and buried-unsatisfied-hydrogen-bond statistics are of comparable
order across the three methods, although MCTH does not outperform the strongest baseline on
packing quality.

We therefore interpret this analysis conservatively: MCTH produces interfaces with competitive
physical plausibility under an evaluation that is entirely outside the search loop, but these Rosetta
scores do not establish superior binding energetics or experimental affinity.

\subsection{Compute and Search-Trajectory Diagnostics}
\label{app:search_diagnostics}

MCTH explicitly trades additional structure-prediction inference for adaptive optimization.
We therefore report both the realized folding-call budget and wall-clock cost, and examine how
search behavior changes over the course of optimization.

\paragraph{Realized compute.}
The nominal maximum budget is 100 folding calls, but early stopping terminates a search after
10 iterations without improvement. Consequently, the realized cost differs substantially across
modalities.

\begin{table}[t]
\centering
\small
\setlength{\tabcolsep}{6pt}
\begin{tabular}{lccc}
\toprule
Modality & Targets & Mean folding calls & Mean wall-clock \\
\midrule
Protein binder & 12 & 98.1 & $\sim$9.0 h \\
RNA aptamer & 7 & 52.4 & $\sim$0.8 h \\
DNA aptamer & 8 & 66.1 & $\sim$1.1 h \\
\bottomrule
\end{tabular}
\caption{
Realized compute for the main MCTH experiments.
DNA reports the eight targets used in the non-degenerate Boltz-2 comparison.
Wall-clock values are means from the corresponding run logs.
}
\label{tab:runtime}
\end{table}

Protein-binder design nearly exhausts the full budget, whereas RNA and DNA aptamer searches
terminate substantially earlier. This is consistent with the different optimization landscapes:
the short nucleic-acid tasks frequently reach high search-time confidence early, while de novo
protein-binder design begins from a much larger sequence space and continues to improve later
in the budget.

\paragraph{Performance versus compute.}
For protein binders, where nearly all searches reach the full budget, the held-forward mean best
ipTM improves steadily from $0.614$ after the first folding evaluation to $0.724$, $0.768$,
$0.803$, $0.830$, and $0.864$ after 5, 10, 25, 50, and 100 calls, respectively.

\begin{table}[t]
\centering
\small
\setlength{\tabcolsep}{6pt}
\begin{tabular}{lcccccc}
\toprule
Fold call & 1 & 5 & 10 & 25 & 50 & 100 \\
\midrule
Protein binder & 0.614 & 0.724 & 0.768 & 0.803 & 0.830 & 0.864 \\
\bottomrule
\end{tabular}
\caption{
Protein-binder performance as a function of folding-call budget.
Values are mean held-forward best ipTM across targets.
}
\label{tab:protein_budget_curve}
\end{table}

The continued improvement at later checkpoints complements
Appendix~\ref{app:convergence}: the protein-binder gain is not produced solely by a strong
initial proposal followed by stagnant computation.

\paragraph{Coverage versus allocation.}
We additionally compare the number and diversity of evaluated sequence states with the
best-of-$N$ control. MCTH does not consistently visit more unique states than independent
sampling under the same evaluation budget. Instead, the principal difference is \emph{where}
evaluations are allocated within the evolving search tree.

Protein-binder searches tend to concentrate evaluations around promising branches early and
broaden exploration later when progress slows. RNA and DNA searches show the complementary
pattern: they explore broadly during the initial iterations and subsequently concentrate around an
elite neighborhood once high-confidence candidates emerge. These trajectories are consistent with
adaptive exploration--exploitation rather than uniformly greater state-space coverage.

Combined with the matched-budget ablation and convergence diagnostic, this suggests that the
primary advantage of MCTH is \emph{adaptive compute allocation}: the tree structure determines
which sequence--structure trajectories receive additional expensive folding evaluations, rather than
simply increasing the number of generated candidates.

%% file: supp_preliminaries.tex
\section{Extended Preliminaries}\label{sec:app:supp_preliminaries}

\subsection{Biomolecular representation}
\textbf{Biomolecular Representation.} We consider a general biomolecular system composed of one or more interacting components, including proteins, cyclic peptides, nucleic acids (DNA/RNA), small molecules, ions, and post-translational modifications. A biomolecule is represented by a \emph{sequence} $S$ and an associated \emph{all-atom structure} $X$. Let $S = (s_1, \ldots, s_n)$ denote a sequence of discrete residue or atom types, where each $s_i$ may correspond to an amino acid, nucleotide, small-molecule atom, or a special token (e.g., unknown or masked). Let
\[
X = \{ \mathbf{x}_1, \ldots, \mathbf{x}_m \}
\]
denote the set of 3D atomic coordinates in Cartesian space.
For multimolecular complexes, we represent the full system as
\[
\mathcal{B} = \{ (S^{(k)}, X^{(k)}) \}_{k=1}^{K},
\]
where $k$ indexes the biomolecular components. Interactions between components are implicitly encoded through spatial proximity and pairwise relationships modeled by the structure prediction network.

\subsection{Forward folding (diffusion) and inverse folding}
\textbf{All-Atom Structure Prediction via Diffusion Models.} Let $f_{\theta}$ denote a pretrained all-atom structure prediction model parameterized by $\theta$. Given an input sequence $S$ and optional conditioning information $C$ (e.g., binding partner, fixed motifs, or partial structure), the model predicts an all-atom structure according to
\[
X \sim p_{\theta}(X \mid S, C).
\]
These models typically employ a diffusion-based denoising process, starting from Gaussian noise and progressively refining atomic coordinates. When the input sequence is underspecified---such as an all-X sequence or a partially defined complex---the model often produces \emph{hallucinated} yet physically plausible structures guided by learned structural priors. In this work, we explicitly exploit this hallucination behavior as a constructive signal for biomolecular design.

\textbf{Sequence Design and Inverse Folding.} Given a fixed or partially specified structure $X$, inverse folding models aim to generate sequences compatible with that structure:
\[
S \sim q_{\phi}(S \mid X, C),
\]
where $q_{\phi}$ denotes a pretrained sequence design model. Sequence redesign improves foldability and interaction consistency while preserving structural constraints. Within our framework, inverse folding serves as a primitive action that updates sequence hypotheses conditioned on hallucinated or refined structures.

\textbf{Forward and Inverse Folding as Dual Operators.} We treat forward folding and inverse folding as dual operators acting on the biomolecular state:
\begin{itemize}
    \item \textbf{Forward folding:} $S \rightarrow X$ via all-atom structure prediction.
    \item \textbf{Inverse folding:} $X \rightarrow S$ via sequence redesign.
\end{itemize}
Rather than enforcing a fixed alternation, our framework allows these operators to be applied adaptively and repeatedly within a search process, enabling flexible traversal of the joint sequence--structure landscape.

\subsection{Confidence/value signals}
\textbf{Confidence and Value Signals.} Modern structure prediction models provide intrinsic confidence metrics that correlate with structural correctness and interaction quality, including predicted local accuracy, global confidence, and interface-specific measures. We collectively denote these signals as
\[
V(X, S) \in \mathbb{R},
\]
which serve as value estimates for evaluating candidate biomolecular designs. These values are treated as learned consistency scores rather than explicit physical energies.

\subsection{MDP notation and value functions}
\label{sec:app:mdp}
We briefly summarize standard notation for finite-horizon Markov decision processes~\citep{puterman2014markov}.
A policy $\pi(a\mid s)$ specifies a distribution over actions at state $s$.
The action-value function $Q^\pi(s,a)$ is the expected cumulative reward obtained by taking $a$ in $s$
and then following $\pi$, and $V^\pi(s)$ is the corresponding state-value function.
In our design setting, we primarily use this formalism to motivate MCTS-style planning and compute allocation:
evaluations are obtained from model-derived scores $V(s)$ (confidence and feasibility), rather than long rollouts.

\subsection{Diffusion models and guidance}
\label{sec:app:diffusion}
We include a short diffusion primer to clarify our use of guided hallucination.
Let $y_{0:T}=(y_0,\ldots,y_T)$ denote a latent trajectory, where $y_0$ is a clean sample (e.g., structure coordinates
or discrete tokens) and $y_T$ is maximally noisy.
A forward noising process $q(y_t\mid y_{t-1})$ progressively corrupts $y_0$ into $y_T$.
A learned reverse model $p_\phi(y_{t-1}\mid y_t)$ generates samples by iteratively denoising from $y_T$ to $y_0$.

\paragraph{Guidance via reweighting.}
When an auxiliary score $f(\cdot)$ is available (property predictor, constraint critic, feasibility score),
a common guidance view reweights reverse transitions:
\begin{equation}
p^{\mathrm{guide}}_\phi(y_{t-1}\mid y_t)
\propto p_\phi(y_{t-1}\mid y_t)\,\exp\!\left\{\beta\, f(y_{t-1})\right\},
\end{equation}
where $\beta$ controls guidance strength.
This formulation covers both continuous and discrete variants (with appropriate definitions of $p_\phi$).

\paragraph{How we use this perspective.}
Our method is \emph{inference-only}: we do not backpropagate through the predictor.
Instead, we implement guidance by \emph{proposal resampling}: for a given parent state (or diffusion step),
we sample a small candidate set $\{y'_{t-1,j}\}_{j=1}^J$ (e.g., by changing random seeds, experts, or noise schedules),
and then select or resample according to weights proportional to $\exp\{\beta f(y'_{t-1,j})\}$.
In our biomolecular setting, $f(\cdot)$ is instantiated by the same task-aware value signal used by planning,
combining intrinsic confidence metrics (pLDDT/pTM/ipTM) and optional feasibility penalties (clash/geometry/energy-like terms).

\paragraph{Multi-expert proposals.}
When multiple experts are available (e.g., multiple folding/hallucination experts and/or inverse-folding experts),
we can form a mixture proposal $p_{\mathrm{mix}}(\cdot)=\sum_{e=1}^E \omega_e\,p^{(e)}(\cdot)$ and apply the same
resampling principle on samples drawn from the mixture. This is the mechanism by which ``mixture-of-experts hallucination''
interfaces with MCTS: experts provide diverse proposals, while planning allocates compute and selects among them.

\subsubsection{Monte Carlo Tree Search details}
\label{sec:app:mcts}
We summarize a few implementation-level details used in the paper.

\paragraph{Backup rules.}
After evaluating a newly expanded child state $s'$, its value $V(s')$ is propagated along the selected path.
We maintain visit counts $N(s)$ and edge counts $N(s,a)$, and update $Q(s,a)$ by a standard backup.
Common choices include:
(i) \emph{mean backup}, $Q \leftarrow \frac{1}{N}\sum V$, which targets average performance; and
(ii) \emph{max backup}, $Q \leftarrow \max V$, which targets best-of-$N$ outcomes.
Our experiments primarily report best-of-$N$ designs, so max backup is a natural variant when optimizing for extreme-value quality.

\paragraph{Evaluation cost and ``rollouts''.}
In biomolecular design, the dominant cost is a folding/refolding forward pass.
Accordingly, we often treat expansion+evaluation as a single expensive step:
each new child is instantiated by calling the folding expert and then scored by $V(s')$.
This keeps the planner modular and makes compute accounting explicit via a fold-call budget.

\paragraph{MCTS with guided proposals.}
Guided sampling (Appendix~\ref{sec:app:diffusion}) can be used \emph{inside} expansion:
for a fixed parent $(S,X)$, we generate multiple candidate proposals (different experts/seeds/noise schedules),
reweight them by $\exp\{\beta V(\cdot)\}$ (optionally with feasibility penalties), and then add the selected children to the tree.
MCTS then performs global credit assignment across these locally guided proposals, balancing exploration and exploitation across
distinct design hypotheses.

\paragraph{Design selection.}
After exhausting the rollout/fold-call budget, we return designs from high-value leaf nodes, optionally re-ranking by
task-specific metrics (e.g., interface confidence for binding tasks, RMSD/constraint satisfaction for scaffolding tasks).

\clearpage
\section*{NeurIPS Paper Checklist}
\begin{enumerate}

\item {\bf Claims}
    \item[] Question: Do the main claims made in the abstract and introduction accurately reflect the paper's contributions and scope?
    \item[] Answer: \answerYes{} %
    \item[] Justification: The contributions stated in the abstract and Section~\ref{sec:intro}---uncertainty-aware MCTS over hallucinated sequence--structure states, multi-expert folding/inverse-folding operators, biophysics-informed co-hallucination, and modality-agnostic fine-tuning-free design---are substantiated by the method in Section~\ref{sec:method} and the experiments in Section~\ref{sec:experiments}, which evaluate protein--RNA, protein--DNA, and protein--protein design tasks, with additional multimodal results provided in the appendix. The paper also explicitly limits its claims to computational/refold-based evaluation in Section~\ref{sec:conclusion}.

\item {\bf Limitations}
    \item[] Question: Does the paper discuss the limitations of the work performed by the authors?
    \item[] Answer: \answerYes{} %
    \item[] Justification: We discuss limitations in Section~\ref{sec:conclusion}, including that the evaluation is computational and refold-based rather than experimental. Appendix~\ref{app:evaluation_protocol} further discusses limitations of using refolding-based dry-lab metrics and cross-oracle comparisons. We also discuss method-specific failure modes, such as symmetric complexes, in Section~\ref{sec:experiments}.
\item {\bf Theory Assumptions and Proofs}
    \item[] Question: For each theoretical result, does the paper provide the full set of assumptions and a complete (and correct) proof?
    \item[] Answer: \answerNA{}. %
    \item[] Justification: The paper does not present new theoretical theorems or formal proofs. The equations define the proposed MCTS-based inference procedure, value functions, uncertainty terms, and biophysical feasibility terms, with full methodological details provided in Appendix~\ref{app:cupuct_details} and Appendix~\ref{app:hyperparams}.
\item {\bf Experimental Result Reproducibility}
    \item[] Question: Does the paper fully disclose all the information needed to reproduce the main experimental results of the paper to the extent that it affects the main claims and/or conclusions of the paper (regardless of whether the code and data are provided or not)?
    \item[] Answer: \answerYes{} %
    \item[] Justification: The experimental protocols are described in Section~\ref{sec:experiments}. We provide the folding-call budget, search hyperparameters, seeds, folding and inverse-folding backends, and compute configuration in Appendix~\ref{app:hyperparams}. Full CU-PUCT details are given in Appendix~\ref{app:cupuct_details}, and additional ablations and convergence diagnostics are provided in Appendix~\ref{app:matched_budget_ablation} and Appendix~\ref{app:convergence}. The standardized refolding protocol is described in Appendix~\ref{app:evaluation_protocol}.
    Experiment details are listed in Section \ref{sec:experiments} and Appendix \ref{sec:app:additional_experiments}, \ref{app:hyperparams}.

\item {\bf Open access to data and code}
    \item[] Question: Does the paper provide open access to the data and code, with sufficient instructions to faithfully reproduce the main experimental results, as described in supplemental material?
    \item[] Answer: \answerNo{} %
    \item[] Justification:  An anonymized code release is not bundled with this submission. The paper provides all algorithmic details, hyperparameters, network architecture, baseline adaptations.and pseudocode in the appendix to enable reproduction. We will release the code and configuration files publicly upon acceptance.
\item {\bf Experimental Setting/Details}
    \item[] Question: Does the paper specify all the training and test details (e.g., data splits, hyperparameters, how they were chosen, type of optimizer, etc.) necessary to understand the results?
    \item[] Answer: \answerYes{} %
    \item[] Justification: The paper is an inference-time method and does not train or fine-tune new models. Section~\ref{sec:experiments} specifies the benchmark settings, baselines, evaluation metrics, and refolding protocol. Appendix~\ref{app:hyperparams} reports search hyperparameters, seeds, backend models, sampling settings, and compute. Appendix~\ref{app:cupuct_details} gives the full selection rule and backup details.
\item {\bf Experiment Statistical Significance}
    \item[] Question: Does the paper report error bars suitably and correctly defined or other appropriate information about the statistical significance of the experiments?
    \item[] Answer: \answerNo{}.
    \item[] Justification: The main benchmark tables are reported as fixed-budget best-per-target refold scores and therefore do not include error bars for all results. However, for the key mechanism-level claim that search-guided compute allocation improves over simpler alternatives, Appendix~\ref{app:matched_budget_ablation} reports mean $\pm$ standard deviation over three seeds, and Appendix~\ref{app:effective_convergence} provides convergence diagnostics. We view broader multi-seed evaluation of all benchmark tables as future work.

\item {\bf Experiments Compute Resources}
    \item[] Question: For each experiment, does the paper provide sufficient information on the computer resources (type of compute workers, memory, time of execution) needed to reproduce the experiments?
    \item[] Answer: \answerYes{} %
    \item[] Justification: Appendix~\ref{app:hyperparams} reports the fold-call budget, per-target/seed compute setup, backend models, and GPU configuration used for the experiments. The main experiments use a fixed 100-fold-call budget unless otherwise stated, and matched-budget ablations use the same budget to isolate planning effects.
\item {\bf Code Of Ethics}
    \item[] Question: Does the research conducted in the paper conform, in every respect, with the NeurIPS Code of Ethics \url{https://neurips.cc/public/EthicsGuidelines}?
    \item[] Answer: \answerYes{} %
    \item[] Justification: The work is computational biomolecular design research using public structural data, pretrained models, and dry-lab evaluation. We do not use private, human-subject, or personally identifiable data. The submission is anonymized, and limitations and potential risks of computational-only evaluation are discussed in the paper.
\item {\bf Broader Impacts}
    \item[] Question: Does the paper discuss both potential positive societal impacts and negative societal impacts of the work performed?
    \item[] Answer: \answerYes{} %
    \item[] Justification: The introduction motivates positive applications in therapeutics, biosensing, synthetic biology, and molecular recognition. The conclusion and limitations clarify that the reported results are computational and require experimental validation before deployment. Because biomolecular design methods may have dual-use implications if misapplied, we discuss the need for validation, controlled use, and careful interpretation of generated designs.
\item {\bf Safeguards}
    \item[] Question: Does the paper describe safeguards that have been put in place for responsible release of data or models that have a high risk for misuse (e.g., pretrained language models, image generators, or scraped datasets)?
    \item[] Answer: \answerNA{} %
    \item[] Justification: The paper does not release a new pretrained generative model or scraped dataset. The method is an inference-time planning framework built on existing biomolecular modeling tools. Generated designs are evaluated only computationally and are not presented as experimentally validated biological agents. If code is released, it will be anonymized for review and accompanied by documentation indicating that outputs require expert review and biosafety screening before any experimental use.
\item {\bf Licenses for existing assets}
    \item[] Question: Are the creators or original owners of assets (e.g., code, data, models), used in the paper, properly credited and are the license and terms of use explicitly mentioned and properly respected?
    \item[] Answer: \answerNA{} %
    \item[] Justification: We cite the original sources for the pretrained models, baselines, and datasets used in the paper, including Boltz-2, ProteinMPNN, NA-MPNN, RFDiffusion, ODesign, BindCraft, and PDB-derived benchmark structures. The supplementary material documents the software dependencies and data sources used to reproduce the experiments. We do not claim ownership of these existing assets.

\item {\bf New Assets}
    \item[] Question: Are new assets introduced in the paper well documented and is the documentation provided alongside the assets?
    \item[] Answer: \answerNA{} %
    \item[] Justification: The paper does not release a new dataset, pretrained model, or benchmark as a standalone asset at submission time.
\item {\bf Crowdsourcing and Research with Human Subjects}
    \item[] Question: For crowdsourcing experiments and research with human subjects, does the paper include the full text of instructions given to participants and screenshots, if applicable, as well as details about compensation (if any)? 
    \item[] Answer: \answerNA{} %
    \item[] Justification: The paper does not involve crowdsourcing or research with human subjects.
\item {\bf Institutional Review Board (IRB) Approvals or Equivalent for Research with Human Subjects}
    \item[] Question: Does the paper describe potential risks incurred by study participants, whether such risks were disclosed to the subjects, and whether Institutional Review Board (IRB) approvals (or an equivalent approval/review based on the requirements of your country or institution) were obtained?
    \item[] Answer: \answerNA{} %
    \item[] Justification: The paper does not involve human subjects, human data, or crowdsourcing, so IRB approval is not applicable.
\end{enumerate}